\documentclass[12pt]{article}
\usepackage{epsfig}

\usepackage{exscale}

\font\fourteenbf=cmbx12 scaled\magstep1

\def\a0size{6}

\newcommand{\lsi}{\raise0.3ex\hbox{$<$\kern-0.75em\raise-1.1ex\hbox{$\sim$}}}
\newcommand{\gsi}{\raise0.3ex\hbox{$>$\kern-0.75em\raise-1.1ex\hbox{$\sim$}}}
\newcommand{\lsim}{\mathop{\lsi}}

\newcommand{\chat}{\hat{C}}
\newcommand{\deltav}{\delta^{(S^{d-1})}}
\newcommand{\gvec}[1]{\mbox{ \boldmath  $#1$}}
\newcommand{\ghat}{\hat{G}}

\renewcommand{\vec}[1]{{\bf #1}}

\newcommand{\intv}[1]{\int_{\vec v _{#1}}}
\newcommand{\half}{\mbox{$\frac12$}}

\newcommand{\im}{{\rm Im}}
\newcommand{\re}{{\rm Re}}
\newcommand{\vbra}{\langle}
\newcommand{\vhat}{\hat{v}}
\newcommand{\vket}{\rangle}

\newcommand{\mmdebye}{m^2_{\rm D}}

\newcommand{\tr}{_{\rm t}}

\newcommand{\mref}[1]{(\ref{#1})}

\begin{document}

\setlength{\baselineskip}{0.6cm}
\newcommand{\figysize}{16.0cm}
\newcommand{\figtopspace}{\vspace*{-1.5cm}}
\newcommand{\figbottomspace}{\vspace*{-5.0cm}}
  
\renewcommand{\theequation}{\thesection.\arabic{equation}}

\newcounter{saveeqn}

\newcommand{\alphaeqn}{\refstepcounter{equation}\setcounter{saveeqn}{\value{equation}}%
\setcounter{equation}{0}%
\renewcommand{\theequation}{%
        \mbox{\thesection.\arabic{saveeqn}\alph{equation}}}}%

\newcommand{\reseteqn}{\setcounter{equation}{\value{saveeqn}}%
\renewcommand{\theequation}{\thesection.\arabic{equation}}}

\begin{titlepage}
\begin{flushright}
BI-TP 2002/08
\\
\end{flushright}
\begin{centering}
\vfill

{\fourteenbf \centerline{ 
Perturbative and non-perturbative aspects 
}
\vskip 2mm \centerline{ 
of  the 
non-abelian Boltzmann-Langevin equation
 }}

\vspace{1cm}
 

Dietrich B\"odeker \footnote{e-mail: bodeker@physik.uni-bielefeld.de}

\vspace{.6cm} { \em 
Fakult\"at f\"ur Physik, Universit\"at Bielefeld, D-33615 Bielefeld
}

\vspace{2cm}
 
{\bf Abstract}

\vspace{0.5cm}

\end{centering}

\noindent
We study the Boltzmann-Langevin equation which describes the 
dynamics of hot Yang-Mills fields with typical momenta
of order of the magnetic screening scale $ g  ^ 2 T $. 
It is transformed into a path integral and Feynman rules are obtained.
We find that the leading log Langevin equation can be systematically
improved in a well behaved expansion in $ \log(1/g) ^{-1}  $.
The
result by Arnold and Yaffe that the leading log Langevin
equation is still valid at next-to-leading-log order 
is confirmed. We also confirm their result for the next-to-leading-log
damping coefficient, or color conductivity,
which is shown to be gauge fixing independent for a
certain class of gauges. The frequency scale
$ g ^ 2 T $ does not contribute to this result, but it {\em does} contribute,
by power counting, to the transverse gauge field propagator.
Going beyond a perturbative
 expansion we find 
1-loop ultraviolet divergences which cannot
be removed by renormalizing the parameters in the Boltzmann-Langevin equation.

\vspace{0.5cm}\noindent


\vspace{0.3cm}\noindent
 
\vfill \vfill
\noindent
 
\end{titlepage}
 
\section{Introduction}
\label{sc:introduction} 

We consider hot non-abelian gauge fields at a temperature $ T $ sufficiently
large so that the 
the running 
coupling $ g=g(T) $ is small. Even then gauge fields 
with typical momenta of order 
$  g ^ 2 T $,  the so called magnetic screening scale, are strongly coupled
\cite{linde}.
Despite this, 
physical quantities have an expansion in powers of $ g $ times possible
logarithms of $ g $. Only the 
coefficients in this expansion cannot be calculated perturbatively when
they receive contributions from magnetic scale momenta. 
For example, the order $ g ^ 6 $ contribution to the free energy comes with
a non-perturbative coefficient. This coefficient  can be calculated 
by a lattice simulation of 3-dimensional pure Yang-Mills theory, which was 
shown by using dimensional reduction \cite{braaten}. This technique can only
be applied to static quantities, like the free energy or equal time 
correlation  functions. 

In this paper we consider dynamical quantities, which are determined by unequal
(real) 
time correlation functions.  While the free energy is sensitive to the 
magnetic scale 
only at relatively high order in the weak coupling expansion, there are
dynamical quantities which are determined by this scale at leading order. An
important example is the Chern-Simons diffusion rate in unbroken non-abelian
gauge theories, often referred to
as the hot sphaleron rate. In the standard electroweak theory it determines the
rate for anomalous baryon number violation above the electroweak phase
transition or cross-over temperature $ T _{\rm c} \sim 100 $ GeV 
\footnote{Close to $ T _{\rm c} \sim 100 $ one must take the Higgs field
into account \cite{moore:higgs}
unless the transition is strongly first order.}.
It is a crucial ingredient in scenarios which try to explain the baryon
asymmetry of the universe, e.g., through leptogenesis \cite{wb:review} 
or electroweak baryogenesis
 \cite{rubakov}.  In QCD Chern-Simons diffusion leads
to non-conservation of quark helicity even in the chiral limit \cite{mottola}.
Another physical quantity which is sensitive to the magnetic screening scale at
leading order is the friction experienced by an electroweak phase transition 
bubble wall \cite{moore:friction}. 

Euclidean lattice simulations are of limited use for calculating real time
observables~
\footnote{Recently
methods have been developed for obtaining real time correlation functions
from Euclidean lattice simulations using the so called maximal entropy method 
\cite{mom}. However, this cannot be applied to Chern-Simons diffusion 
which is exponentially
suppressed in Euclidean time.}, and real time simulations of quantum
field theory are not possible. 
However, even for dynamical quantities one can use perturbation
theory to integrate out short distance modes to obtain an effective
theory for the non-perturbative long-distance dynamics.  The first
step is to integrate out ``hard'' physics associated with virtual
momenta of order $ T $. At leading order in the gradient expansion one
obtains the so called hard thermal loop effective theory \cite{htl}.
The remaining degrees of freedom are gauge fields with characteristic
wave vectors  $ \vec k \lsim  g T $. Their occupation number $ n(|\vec k|) 
= 1/(e ^{|\vec k|/T } -1 ) \simeq T/|\vec k|$ is large which means that 
they can be treated as classical fields.  
This opens the  possibility 
for a non-perturbative treatment because classical field theories can 
be treated
on the lattice in real time \cite{grigoriev}. 
The hard thermal loop effective theory was used
for non-perturbative lattice simulations \cite{moore:htl}-\cite{bmr}
which, however, do not have 
a continuum limit \cite{bms}. 

One can, however,  go further using perturbation theory. 
Integrating out all frequency scales and spatial momenta 
\footnote{Since Lorentz invariance is lost at finite temperature,
frequencies and spatial momenta have to be treated
separately.} 
 larger than $
g ^ 2 T $ one finds that at leading order in $ g $  the dynamics of
gauge fields $ A ^ \mu  $ is described by the Boltzmann-Langevin 
equation \cite{local}
\begin{eqnarray}
        (C + v \cdot D) \widetilde{W}  
        + \frac{ d}{m^2_{\rm D} }(c_1 + \vec v \cdot \vec D)
        \vec v \cdot \vec D \times \vec B 
        = \vec v \cdot \vec E + \xi
        \label{boltzmann}
        ,
\end{eqnarray}
which is a classical field equation for $ A $ and $ \widetilde{W} $. 
$ D ^  \mu _ {ab} = \delta _ {ab} \partial ^  \mu - g f _ {abc} A ^ 
\mu _  c $ is the covariant derivative in the adjoint representation, $
E ^ i = F ^{i0} $ and $ B ^ i = -\half \epsilon ^{ijk} F ^{jk} $ are
the non-abelian electric and magnetic fields. $ m_{\rm D} $ is the Debye or
electric screening mass which is of order $ g T $, and $ d $ denotes the 
number of spatial dimensions. We choose $ d = 3 - \epsilon  $ to 
regularize ultraviolet and infrared divergences.
$ \widetilde{W} (x, \vec v) $ transforms under the adjoint representation
and
represents color charge fluctuations of  plasma particles  with
momenta of order $ T $ and velocity $ \vec v $.  
We neglect fermion masses. Then  these
particles move at the speed of light so that   $ \vec v ^ 2 = 1 $.
The 4-vector  $ v $ is defined as $ v ^ \mu  = (1, \vec v) $.   
$ C $  is a linear collision operator of order $ g ^ 2 T $  
and depends logarithmically on the cutoff separating the scales 
$ g T $ and $ g ^ 2 T $. It is 
local in space and acts only on the velocity variable,
\begin{eqnarray} 
  C f (\vec v) \equiv \int _ {\vec v '} C(\vec v, \vec v ') f(\vec v')
  .
\end{eqnarray} 
Here $ \int _ {\vec v}(\cdots ) 
 \equiv \Omega  ^{-1} \int d ^ {d - 1} v   (\cdots )$ is the 
normalized integral
over the $ d - 1 $ dimensional unit sphere, $ \Omega = 4 \pi  $ for $
d = 3 $. 
$ C $ commutes with rotations of $ \vec v $. Therefore it can be
diagonalized by expanding $ W(x,\vec v) $ in spherical harmonics 
\footnote{Or their generalization in $ d \neq 3 $ dimensions.
To keep the terminology simple, we refer to them as spherical harmonics.}
$ Y _{lm}
(\vec v)$,  and the eigenvalues depend only on $ l $. The 
$ l = 0 $ eigenvalue vanishes, and the coefficient $ c _ 1 $ in 
Eq.~(\ref{boltzmann}) is the $ l = 1 $ eigenvalue. 
$ \widetilde{W} $ contains only spherical harmonics with respect $ \vec
v $ with $ l \ge 2 $, i.e.,
\begin{eqnarray}
        \int _{\vec v} v ^ \mu  \widetilde{W} (x, \vec v) = 0
        .
\end{eqnarray} 
$ \xi (x, \vec v) $ is a Gaussian white noise with the 2-point function
\begin{eqnarray}
        \langle \xi _   a(x,\vec v) \xi _  b(x',\vec v ') \rangle =
        \frac{2T}{\mmdebye} C(\vec v, \vec v ') \delta _  {ab} 
        \delta^{d+1}(x
        - x') \label{xicorrelator}
        .
\end{eqnarray}
A non-trivial check of Eq.~(\ref{boltzmann}) 
is to see whether it reproduces the known thermodynamics
of the magnetic scale gauge fields. In Ref.~\cite{local}  it was
indeed shown that Eq.~(\ref{boltzmann}) generates a thermal ensemble with
a Boltzmann weight $ e ^{-H/T } $ where
\begin{eqnarray}
        H_ {\rm }&=& \frac12 \int d^3 x  \left\{
        \vec{B}^2
        + \mmdebye\int_{\vec v}
        \widetilde{W}^2          \right \}
        \label{hamiltonian}
        .
\end{eqnarray}
Thus for equal time correlation functions  $ \widetilde{W} $  acts
like a free field and does not affect the gauge fields. The term
$ \frac{ 1}{2} \vec B ^ 2 $ is  just the Lagrangian of the dimensionally
reduced theory for the scale $ g ^ 2 T $ \cite{braaten}.

In the leading log approximation, neglecting 
terms suppressed by powers of $ [\log (1/g)]^{-1} $, 
Eq.~(\ref{boltzmann}) can be replaced by a much simpler equation 
of motion 
which contains only the gauge fields \cite{ll}, 
\begin{eqnarray}
  \vec D \times \vec B = \gamma  \vec E + \gvec{\zeta   }
  \label{langevin} 
  .
\end{eqnarray}
Here
$ \gvec{\zeta } $ is again a Gaussian white noise which is proportional
to the $ l=1 $ projection of $ \xi  $. The damping coefficient $ \gamma  $,
the color conductivity, is proportional to 
$ T/\log(1/g) $. The only spatial momentum scale in Eq.~(\ref{langevin}) is
$ \vec k \sim g ^ 2 T $. Estimating the LHS as $ \vec k ^ 2 \vec A $ and the
first term on the RHS as $ \gamma  k ^ 0 \vec A$ one can see that
the only frequency scale  in Eq.~(\ref{langevin}) is 
$ k ^ 0 \sim g ^ 4 T$ (modulo logarithms of the coupling).  
The Boltzmann-Langevin equation (\ref{boltzmann}), 
on the other hand,   
contains two frequency scales, $ g ^ 4 T $ and $ g ^ 2 T $, again modulo
logarithms \cite{local}. The question whether the frequency 
scale $ g ^ 2 T $ affects  the non-perturbative
gauge field dynamics has not been addressed so far. 

Eq.~(\ref{langevin}) solves the notorious ultraviolet problems mentioned above.
It is ultraviolet finite and has been used to calculate the hot sphaleron 
rate by Moore \cite{moore:log}. Its obvious disadvantage is that it's only
valid in the logarithmic approximation. 
The Boltzmann equation
which was  obtained by integrating out the momentum scale $
g T $ \cite{ll} 
\footnote{ For a Feynman-diagram interpretation, 
see \cite{ladder,Blaizot:2000}.}
 and which served as a starting point for obtaining
Eqs.~(\ref{langevin}) and (\ref{boltzmann}) is valid to leading order in
$ g $ and all orders in $ [\log(1/g)] ^ 
{-1} $, but it has the same 1-loop ultraviolet
divergences as classical Yang-Mills theory.  One can hope that
Eq.~(\ref{boltzmann}) has a better ultraviolet behavior because it does not
contain propagating gauge field modes which are responsible for these 
divergences.

An equation similar to (\ref{boltzmann}) has been obtained 
by Arnold \cite{arnold:langevin}. 
The difference between Arnold's Langevin equation  and Eq.~(\ref{boltzmann})
is that the former does not contain a time derivative of $ \widetilde{W} $.
Thus in   \cite{arnold:langevin} 
$ \widetilde{W}   $ is not dynamical, but it  is 
fixed  by $ A  $  and $ \xi  $ at  the same instant of time. The only 
 dynamical degrees of freedom are  the gauge fields. 
On first sight this difference appears to
be minor. 
However, if one discretizes time
for solving these equations on the
lattice with temporal lattice spacing $ \Delta  x ^ 0 $  the 
two equations require rather different treatments.
For Eq.~(\ref{boltzmann}) $ A  $ and
$ \widetilde{W}  $ at time $ x ^ 0  + \Delta  x ^ 0 $ and spatial coordinate
$  \vec x  $ are determined by
$ A $, $ \widetilde{W} $, a finite number of their derivatives 
and $ \xi  $, 
all taken at $ x ^ \mu  =( x ^ 0 , \vec x ) $. 
This procedure is exactly the same as the one one would use for solving
Eq.~(\ref{langevin}) \cite{moore:log}. 
For Arnold's Langevin equation,
on the other hand, one first has to solve for $ \widetilde{W} $ at time 
$ x ^ 0 $, so that $ \widetilde{W} ( x ^ 0 ) $ becomes a spatially non-local 
functional of $ A $ and $ \xi  $ at time $ x ^ 0 $.  Then one can determine
$ A ( x ^ 0  + \Delta  x ^ 0 ) $ from the so obtained $ \widetilde{W} (
x ^ 0 ) $ and from $ A ( x ^ 0 ) $ and $ \xi  ( x ^ 0 ) $. 
Since 
$ \widetilde{W} $ is linear in $ \xi   $, Arnold's
Langevin equation can be viewed as a spatially non-local generalization
of Eq.~(\ref{langevin}) in which the color conductivity $ \gamma  $ and
the noise $ \xi  $ are gauge field dependent.    The time discretization
 of such a Langevin equation is ambiguous,
but it was argued in Ref.~\cite{arnold:langevin} that the ambiguity
can be fixed by demanding that the equation yields the
correct thermodynamics of magnetic scale gauge fields. 
In the path integral formulation of this equation (cf. Sec.~\ref{sc:path})
it is necessary
to introduce additional terms in the 
action~\footnote{These terms are rather complicated. To quote the
author of Ref.~\cite{arnold:langevin}, 'they are ugly as sin'.}.
No such complications arise for Eq.~(\ref{boltzmann}).

Arnold and Yaffe used the Langevin equation of \cite{arnold:langevin} to show
that Eq.~(\ref{langevin}) is still valid at next-to-leading logarithmic order
(NLLO), if one uses  the next-to-leading logarithmic
order (NLLO) color conductivity $ \gamma $, which was 
calculated in Ref.~\cite{ay:nllo}. It was argued that the additional terms
in the path integral action mentioned above do not 
contribute at  next-to-leading logarithmic order.

The purpose of this paper is three-fold. First, we want to study the large
momentum behavior of Eq.~(\ref{boltzmann}) to see whether it is renormalizable.
This is necessary if one wants to use Eq.~(\ref{boltzmann}) for
non-perturbative lattice simulations and take the continuum limit. 
Secondly, we would like to understand
whether Eq.~(\ref{boltzmann}) gives the same gauge field dynamics as the
Langevin equation obtained by Arnold \cite{arnold:langevin}.
Finally we calculate the color conductivity at next-to-leading logarithmic
order. We confirm the result obtained by Arnold and Yaffe \cite{ay:nllo}, 
and we
investigate its gauge fixing dependence. 

To address these issues it is
convenient to write the equation of motion (\ref{boltzmann}) 
 as a path integral.  This method is briefly reviewed in
Sec.~\ref{sc:path} for a non-gauge theory. To apply it to 
Eq.~(\ref{boltzmann})  one has to fix the
gauge which is discussed in Sec.~\ref{sc:gaugefixing}. 
The path integral representation and the corresponding Feynman rules
for Eq.~(\ref{boltzmann}) are obtained  in Sec.~\ref{sc:pathbl}.
In Sec.~\ref{sc:large} we study the large momentum
behavior of the various propagators. 
We shall find 1-loop ultraviolet divergences which
cannot be removed by renormalizing the parameters of the action
(Sec.~\ref{sc:renormalizability}).  In Sec.~\ref{sc:nllo} 
we discuss the logarithmic approximation beyond leading order. 
Following Arnold and Yaffe we consider 
Wilson loops in Sec.~\ref{sc:wilson} at next-to-leading
logarithmic order (NLLO) to compute the
NLLO color conductivity. 
 In Sec.~\ref{sc:spatial} we study the effect of the modes
with $ k ^ 0 \sim \vec k \sim g ^ 2 T $ 
on the 
spatial gauge field propagator.  Sec.~\ref{sc:summary} contains a
summary and a discussion.  
In Appendix \ref{ap:ghat} we list some properties
of the operator $ \hat{G} $ defined in Eq.~(\ref{ghat}). 
Appendix \ref{ap:estimates}  describes the power counting for propagators
and vertices used in Sects.~\ref{sc:wilson} and \ref{sc:spatial}. It also
contains the approximated expressions for the propagators which are used in 
Sec.~\ref{sc:wilson}.  
The
frequency integrals needed for the 1-loop selfenergies are listed in
Appendix \ref{ap:frequency}. Appendix~\ref{ap:diagrams}  
contains the results for the
individual 1-loop diagrams contributing to the $ A ^ 0 $ propagator
which gives the dominant contribution to the Wilson loops discussed in 
Sec.~\ref{sc:wilson}.

\noindent
{\bf  Notation} 4-vectors and 3-vectors are denoted by italics and boldface,
respectively. The signature of the metric is $ ( {}+---) $.

\section{Path integral representation of Langevin equations}
\setcounter{equation}0 
\label{sc:path} 

To investigate the properties of a Langevin equation 
it is convenient   to re-write it as a path
integral \cite{zj}, which allows the use of the standard
 methods of quantum field theory.
In this section we 
illustrate this method for a non-gauge theory. 
The  fields are denoted by $ \varphi ^ i $ 
and $ i $ represents all indices
including space and time variables. Similarly, we collectively label the 
components of the equation of motion
with the index $ \alpha  $, 
\begin{eqnarray} 
  {\cal E} ^ \alpha [\varphi ]= \xi ^ \alpha  
  \label{eom}
  .
\end{eqnarray} 
We write the 2-point function of the noise as 
\begin{eqnarray}
  \langle \xi  ^ \alpha  \xi  ^ \beta  \rangle = 2 \tau   
  \delta  ^{\alpha   \beta  } 
  .
\end{eqnarray}
Physical observables  are obtained by averaging over the noise with
a Gaussian weight 
\begin{eqnarray}
  \langle O \rangle = \int [d \xi  ] \exp \left\{ 
    -\frac{ 1}{4 \tau   } \xi ^ \alpha \xi ^ \alpha \right\} 
  O(\varphi  _{\rm solultion})
  ,
\end{eqnarray} 
where $ \varphi  _{\rm solution} $ is the solution to the equation of motion
(\ref{eom}). Now one introduces an integral over $ \varphi  $ via 
\begin{eqnarray}
  1 = \int [d \varphi ] \mbox{det} \left( 
    \frac{ \delta {\cal E ^ \alpha  } }{\delta \varphi ^ i} \right) 
  \delta({\cal E - \xi  } ) 
  \label{eins} 
  .
\end{eqnarray}
The $ \delta $-function in (\ref{eins}) 
is represented by an integral over a Lagrange
multiplier field $ \lambda  $, 
\begin{eqnarray}
  \langle O \rangle = \int [d \varphi \, d \xi \, d \lambda  ] \mbox{det} \left( 
    \frac{ \delta {\cal E ^ \alpha  } }{\delta \varphi ^ i} \right) 
  \exp \left\{- 
    \frac{ 1}{4 \tau  } \xi ^ \alpha \xi ^ \alpha + i \lambda 
  ^ \alpha ({\cal E} ^ \alpha - \xi  ^ \alpha  )  \right\} 
  O[\varphi  ]
  .
\end{eqnarray} 
Finally one integrates over the noise and introduces ghost and anti-ghost
fields $ \eta  $ and $ \bar{\eta  } $ to represent the determinant
which gives
\begin{eqnarray}
   \langle O \rangle = \int [d \varphi \, d \lambda \, d \eta  \, d \bar{\eta  } ]
   e ^{-S} 
  O[\varphi  ]
  .
\end{eqnarray} 
The action 
\begin{eqnarray} 
  S =   
    \tau   \lambda  ^ \alpha  \lambda  ^ \alpha  - i \lambda 
  ^ \alpha {\cal E} ^ \alpha    - \bar{\eta  } ^ \alpha  
  \frac{ \delta {\cal E ^ \alpha  } }{\delta \varphi ^ i}
  \eta  ^ i 
\end{eqnarray} 
is invariant under the BRST-transformation 
\begin{eqnarray}
  \delta  \varphi  ^ i = \bar{\varepsilon  } \eta  ^ i, \quad
  \delta  \bar{\eta } ^ \alpha  = -i \bar{\varepsilon  } \lambda  ^ \alpha 
  , \quad \delta  \lambda  ^ \alpha  = \delta  \eta  ^ i = 0
  \label{brs} 
\end{eqnarray} 
with the anti-commuting parameter $ \bar{\varepsilon }  $. 
This procedure can be applied to gauge theories, but first
one has to fix the gauge.

\section{Gauge fixing}
\setcounter{equation}0 
\label{sc:gaugefixing}

Gauge covariant Langevin equations determine the gauge fields only 
up to a gauge transformation. Thus in order to use   Eq.~(\ref{eins}) 
to introduce a path integral one first has to fix  a gauge.
Zinn-Justin and Zwanziger \cite{zjz}  showed that the gauge condition
\begin{eqnarray} 
         A ^ 0 (t) = v[\vec A (t)]
         \label{zg} 
\end{eqnarray}
does not affect expectation values of gauge invariant operators $
O[\vec A] $. A small variation
\begin{eqnarray}
        v[\vec A] \to v[\vec A] + \epsilon [\vec A]
\end{eqnarray} 
can be compensated by a gauge transformation of $ A $. It is easy to see
that this result also holds for gauge invariant operators which depend
both on the spatial gauge fields $ \vec A $ and on $ A ^ 0 $. 

A special case of (\ref{zg}) is
\begin{eqnarray}
  \kappa A ^ 0 + \nabla \cdot \vec A = 0
  \label{flowgauge} 
  ,
\end{eqnarray} 
 which smoothly interpolates between Coulomb ($ \kappa = 0 $) and
temporal gauge ($ \kappa \to \infty $). 
The free fields in this gauge are related to $ A ^ 0 $ in Coulomb gauge
by
\begin{eqnarray}
        A ^ 0 (k) &=& 
        \frac{\vec k ^ 2}{\kappa}\frac{ i}{k ^ 0 + i \vec k ^ 2/\kappa}
         A ^ 0 _{\rm Coul} (k)
         \nonumber 
         ,
\\
        \hat{\vec k} \cdot \vec A  (k) & = & 
         \frac{ |\vec k| }{k ^ 0 + i \vec k ^ 2/\kappa}
        A ^ 0 _{\rm Coul} (k)
        \quad \mbox{(free fields)}
        \label{eomB1} 
        .
\end{eqnarray} 
With this relation we can obtain the propagators for the gauge
(\ref{flowgauge}) from the Coulomb gauge propagators listed in 
Sec.~\ref{sc:propagators}.  

To obtain a path integral representation one can simply use
the gauge condition (\ref{flowgauge}) to eliminate $ A ^ 0 $ from 
the equations of motion, Eq.~(\ref{boltzmann})  or Eq.~(\ref{langevin}),
and then apply the procedure of Sec.~\ref{sc:path}. Closed ghost loops vanish
in this case, so they do not contribute to amplitudes of the fields $
\varphi  $ and $ \lambda  $. 

A slightly different method was used by Arnold and Yaffe \cite{ay:nllo} who
used temporal $ A ^ 0 =0$ gauge to obtain the path integral representation.  
For the
leading log Langevin equation (\ref{langevin}) this 
leads to a supersymmetric action in the path integral. It is not clear,
however, whether the analogous 
action for Eq.~(\ref{boltzmann})  would be supersymmetric.
Then the  Fadeev-Popov procedure was used in \cite{ay:nllo} to switch to 
Coulomb gauge.

\section{Path integral representation of the Boltzmann-Langevin equation}
\setcounter{equation}0 
\label{sc:pathbl}

We can now write the path integral representation of Eq.~(\ref{boltzmann}) 
as
\begin{eqnarray}
  \int [d \vec A \, d \widetilde{W} \, d \lambda \,  d \eta  
  \, d  \bar{\eta  } ]
  e ^{-S} 
  \label{path} 
\end{eqnarray} 
with the  action 
\begin{eqnarray}
  S &=& \int d ^ 4 x \int _{\vec v}\Bigg\{ 
    \frac{ T}{m^2_{\rm D} } \lambda  C \lambda   -i 
     \lambda  \Bigg[ 
        (C + v \cdot D) \widetilde{W}
\nonumber \\ && {}  \hspace{2cm}
        + \frac{ d}{m^2_{\rm D} }(c_1 + \vec v \cdot \vec D)
        \vec v \cdot \vec D \times \vec B 
        -\vec v \cdot \vec E \Bigg]  
        \Bigg\} 
        + S _{\rm ghost}
        \label{action0} 
        .
\end{eqnarray}
As was mentioned above, closed ghost loops vanish in this theory. 
In this paper we are only interested in amplitudes of the physical
fields $ A $ and $ \widetilde{W} $ and of $ \lambda  $. Therefore we
can ignore the ghost part of the action $  S _{\rm ghost} $ in the following. 

The vertices can be read off from the interaction Lagrangian  
\begin{eqnarray}
        S _ {\rm int} &=& 
        -i g f _ {abc} 
        \int _ {k _ 1, k _ 2, k _ 3}  
        (2\pi)^{d+1}\delta(  k _ 1+ k _ 2+ k _ 3)
\nonumber \\ && 
        \Bigg\{
        \int _{\vec v}
        \lambda  _ a(k _ 1,\vec v)[ A ^  0 _ {b} (k _ 2) - 
        \vec v \cdot \vec A _ {b} (k _ 2)] \widetilde{W} _ c(k _ 3 , \vec v)
\nonumber \\ &&
        {}+ \frac{ 1}{2} \frac{ d}{m^2_{\rm D} } 
        \Bigg[
        \left[ -i c _ 1 \lambda ^{i } _ a ( k  _ 1) - k _ 1 ^ j
        \lambda ^{ij} _ a (k _ 1) \right]
\nonumber \\ &&   \hspace{2cm} \times                
        \left[ (k _ 2 - k _ 1)^ n \delta ^{il} 
        + (k _ 3 - k _ 2)^ i \delta ^{nl} 
        + (k _ 1 - k _ 3)^ l \delta ^{in}
        \right]
\nonumber\\ && \hspace{1cm} {}
        + \lambda ^{i m } _ a (k _ 1) 
        \left[ \delta ^{nm} P \tr ^{il}(\vec k _ 2) \vec k _ 2 ^ 2   
        - \delta ^{lm} P \tr ^{in}(\vec k _ 3) \vec k _ 3 ^ 2
        \right] \Bigg]   
        A ^{l} _ b( k _ 2) A ^{n} _ c( k _ 3) 
\nonumber\\ &&
        {}  - 
        \gvec{\lambda} _ { a} ( k  _ 1) 
        \cdot \vec A _ {b}(k _ 2) A ^{0} _ c(k _ 3)  
        + O(\lambda \vec A ^ 3) \Bigg\}
        \label{blN14.3.0}       
\end{eqnarray}
with 
\begin{eqnarray}
  \int _ k \equiv \int \frac{ d ^{d+1}  k}{(2 \pi ) ^{d+1} }
  .
\end{eqnarray} 
Furthermore, we have introduced the $ 3- $vector $ \gvec \lambda  $ through
\begin{eqnarray}
  \lambda  ^ i \equiv \int _{\vec v } v ^ i \lambda  ( \vec v )
  ,
\end{eqnarray} 
and the traceless tensor
\begin{eqnarray}
  \lambda  ^{ij} \equiv
  \int _{\vec v } \left ( v ^ i v ^ j - \frac{ 1}{d} \delta  ^{ij} \right ) 
  \lambda  ( \vec v )
  ,
\end{eqnarray} 
which represent the $ l = 1 $ and $ l = 2 $ projections of $ \lambda  ( \vec 
v ) $, respectively. 
Finally, the 
transverse projector is
\begin{eqnarray}
        P\tr ^{ij}  (\vec k ) = \delta ^{ij} - \hat{k} ^ i \hat{k} ^ j
        .
\end{eqnarray}
In Eq.~(\ref{blN14.3.0}) 
we have not displayed the $ \lambda  \vec A ^ 3 $ and  $ \lambda  \vec A ^ 4 $
vertices which do not play a role in the following. The only interaction
in Eq.~(\ref{blN14.3.0}) which contains derivatives is the   
$ \lambda  \vec A ^ 2 $ vertex. For power counting we can estimate this vertex
as
\begin{eqnarray}
        (\lambda \vec A \vec A) \sim g \frac{ c _ 1 |\vec k |}{m^2_{\rm D}}
        ,
\end{eqnarray} 
where $ \vec k $ is a typical spatial momentum carried by one of the fields. 
The remaining non-derivative interaction vertices are simply of order $ g $.

\subsection{Propagators}
\label{sc:propagators} 

Obtaining the propagators for Eq.~(\ref{action0}) is tedious but
straightforward.  
To simplify the notation it is useful to think of 
$C $ and $ \vec v $
as  operators acting on the space of functions on the  2-sphere.
These functions are represented by bras and kets.  We define $|\vec
v\vket$ as the eigenvector of the operators $\hat{ v}^i$ ($i=1,\ldots
,d$) with eigenvalues $ v^i$,
\begin{eqnarray}
        \hat{v}^i |\vec v\vket =  v^i |\vec v\vket
\end{eqnarray}
and with  the normalization
\begin{eqnarray}
        \vbra \vec v |\vec v ' \vket
        = \Omega \deltav(\vec v - \vec v ')
\end{eqnarray}
Here $ \deltav $ is the delta-function on the $ d-1 $ dimensional unit sphere,
the surface area of which is denoted by $ \Omega  $. 
Then the unit operator is
\begin{eqnarray}
        1 = \int _{\vec v} |\vec v \vket \vbra \vec v|
        .
\end{eqnarray}
It is also convenient to introduce 
\begin{eqnarray}
        |^{\mu}\vket = \intv{} v^\mu  |\vec v\vket
        .
\end{eqnarray}
These are normalized such that
\begin{eqnarray}
        \vbra ^ 0 |^ 0 \vket = 1, \qquad\vbra ^ i |^ j \vket =\frac{ 1}{d}
        \delta ^{ij} 
        .
\end{eqnarray} 
The action (\ref{action0}) can then be written as 
\begin{eqnarray}
        S &=&  \int d ^ 4 x \Bigg\{ \frac{ T}{m^2_{\rm D} }
        \langle \lambda | \hat{C} | \lambda  \rangle
        -i \langle \lambda| \Bigg[  
        (\hat{C} + \hat{v} \cdot D)| \widetilde{W}   \rangle
      \nonumber \\ && {}\hspace{1cm}
        + \frac{ d}{m^2_{\rm D} } (c _ 1 + \hat{\vec v} \cdot \vec D)
        |^i \rangle D _  j F ^{ji} - |^i \rangle E ^{i} 
        \Bigg] \Bigg \} 
      \label{action}
      .
\end{eqnarray}  
To determine the momentum space propagators one has to invert $ \hat{C}
- i \hat{v} \cdot k $. We denote its inverse by
\begin{eqnarray}
        \ghat(k) \equiv \left[ \chat - i \vhat\cdot k \right] ^{-1}
        \label{ghat}
        .
\end{eqnarray}
For writing propagators involving gauge fields it is convenient
to introduce
\begin{eqnarray}
        G ^{\mu  \nu  } \equiv  \langle ^ \mu   |\hat{G}|^ \nu   \rangle  
\end{eqnarray}
which is symmetric under exchange of $ \mu   $ 
and $ \nu  $. Its spatial components are decomposed into transverse
and longitudinal parts, 
\begin{eqnarray} 
        G ^{ij} 
        = P\tr^{ij} G\tr + \hat{k} ^ i \hat{k} ^ j G _ \ell
        .
\end{eqnarray}
Some properties of $ \hat{G} $ and related functions are listed
in Appendix A.

We give the expressions for propagators in Coulomb
gauge. The propagators for general flow gauge can then be obtained by
using Eq.~(\ref{eomB1}). We denote spatial gauge
fields by wavy, $ A _ 0 $ by dashed, $ \widetilde{W} $ by full, and
the Lagrange multiplier fields $ \lambda $ by dotted lines. 

The $ \lambda \lambda $ propagator vanishes,
\begin{eqnarray}
        \lambda
        \,\, \epsfig{file=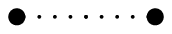,width=1.5cm}\,\, 
        \lambda         &=&        0
        .
\end{eqnarray} 
This is an immediate consequence of the invariance
under the BRST transformation  (\ref{brs}).  Due to
ghost-number conservation we have $ \langle \lambda \bar{\eta } \rangle = 0 $.
Performing a BRST transformation in the path integral 
for $ \langle \lambda \bar{\eta } \rangle = 0 $ then gives $ \langle \lambda
\lambda \rangle =0 $.

The off-diagonal propagators mixing the physical fields
with $ \lambda $ are not invariant under changing sign of the momentum.
Therefore we explicitly display the momentum variables, 
\begin{eqnarray}
        A ^ i (k)
        \,\, \epsfig{file=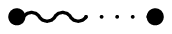,width=1.5cm}\,\, 
        \lambda (-k, \vec v)
        &=& 
        i \langle  S ^ i (k) |\vec v' \rangle 
        ,
\end{eqnarray} 
\begin{eqnarray} 
        A ^  0 (k)
        \,\, \epsfig{file=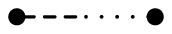,width=1.5cm}\,\, 
        \lambda (-k, \vec v)
        &=& 
        i \langle  S ^ 0 (k) |\vec v \rangle 
        .
\end{eqnarray} 
The expressions
\begin{eqnarray}
        \langle  S ^ i (k) | \equiv m^2_{\rm D}  
        \widetilde{\Delta}(k) 
        P\tr ^{ij}(\vec k)
        \langle ^ j |
        \ghat(k) 
        \label{si} 
        ,
\end{eqnarray}
with
\begin{eqnarray} 
        \widetilde{\Delta } (k) \equiv \left [1 + i d k_0 G\tr(k)\right]^ {-1}
        \Delta (k)
        ,
\end{eqnarray}
\begin{eqnarray}
        \Delta(k) \equiv \frac{ 1}{\vec k ^ 2 - i \mmdebye k _ 0
                G\tr(k)[1 + i d k_0 G\tr(k)]^{ ^{-1}}}
        \label{Delta}
\end{eqnarray} 
and
\begin{eqnarray}
        \langle  S ^ 0 (k) | \equiv   
        \frac{-i   }{ \vec k ^2 G _ \ell (k)} 
        k^i\langle ^{i} | \ghat (k) (1 - \hat{P}_0 ) 
        \label{s0} 
\end{eqnarray} 
will also appear in the $ A ^ \mu - \widetilde{W} $ propagators below. 
The $ \widetilde{W} - \lambda $ propagator can be written as
\begin{eqnarray} 
        \widetilde{W} (k, \vec v)
        \,\, \epsfig{file=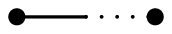,width=1.5cm}\,\, 
\lambda (-k,\vec v') \rangle 
        &=&   
        i 
        \langle \vec v | 
        \left(1 - \hat{P}_0 - \hat{P}_1 \right)
        \hat{S} (k) 
        \left(1 - \hat{P}_0 \right)
        |\vec v ' \rangle
        .
\end{eqnarray} 
$ \hat{P} _ 0$ and $ \hat{P} _ 1 $ are the projectors onto the $ l=0 $ and
$ l = 1 $ sectors,
\begin{eqnarray}
        \hat{P} _ 0   \equiv  |^ 0 \rangle\langle ^ 0  |, 
        \qquad 
        \hat{P}_1 \equiv d |^ i \rangle\langle ^ i  |
\end{eqnarray} 
and
\begin{eqnarray}
        \hat{S}(k)
        & \equiv &
        \ghat(k) - \frac{ 1 }{G _ \ell(k)}
        \ghat(k)|^i\rangle\hat{k}^i\hat{k}^j\langle^j|\ghat(k)
\nonumber \\
        &&{}\hspace{-1cm}
        - \frac{ 1 }{G \tr(k)}
        \left [ 1 - \vec k ^ 2 \widetilde{\Delta } (k) \right]
        \ghat(k)|^i\rangle P\tr^{ij}(\vec k)\langle^j|\ghat(k)
        \label{shat} 
        .
\end{eqnarray}
Note that all propagators which mix $ \lambda $ with the "matter" fields
$ A ^ \mu $ and $ \widetilde{W} $ have poles only in the lower half of
the complex $ k ^ 0 $ plane. As a consequence, loops which contain only
such propagators vanish. 

To obtain the propagators for the physical fields, one uses 
\begin{eqnarray}
        \ghat(k)\chat \ghat(-k) &=& \frac12
        \left[ \ghat(k) + \ghat(-k) \right] 
        \label{trick1}
        ,
\end{eqnarray}
and
\begin{eqnarray}
        \left[ G\tr(k)+ G\tr(-k) \right] 
        \widetilde{\Delta } (k)\widetilde{\Delta } (-k)
        = \frac{ 1}{\mmdebye} \frac{ -i}{k_0} 
        \left[ {\Delta } (k)-{\Delta } (-k)\right] 
        \label{trick2}
        .
\end{eqnarray}
After some algebra one finds that the gauge field propagators take
the form
\begin{eqnarray}
        A ^ i
        \,\, \epsfig{file=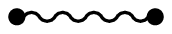,width=1.5cm}\,\, 
        A ^ j    &=& 
        T P\tr ^{ij}
         \frac{- i}{k ^ 0} \Delta(k) + (k\to -k) 
        \label{tree_aa}
        ,
\\
        A ^ 0
        \,\, \epsfig{file=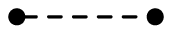,width=1.5cm}\,\, 
        A ^ 0
        &=& 
        \frac{ T}{\mmdebye}     
        \frac{ 1}{\vec k ^ 2 G _ \ell (k)}  + (k\to -k) 
        \label{tree_a0a0}
        .
\end{eqnarray}  
Note that
\begin{eqnarray}
  \Delta ^ *(k) = \Delta (-k)
  .
\end{eqnarray} 
The off-diagonal propagators are
\begin{eqnarray} 
        \widetilde{W} 
        \,\, \epsfig{file=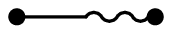,width=1.5cm}\,\, 
     A ^ i 
        &=& 
\frac{ T}{m^2_{\rm D} }
        \langle \vec v | (1 - \hat{P}_1 )  |S ^ i  (k) \rangle 
          + (k\to -k) 
          ,
\\
        \widetilde{W} 
        \,\, \epsfig{file=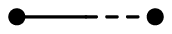,width=1.5cm}\,\, 
     A ^ 0
        &=& 
\frac{ T}{m^2_{\rm D} }
        \langle \vec v | (1 - \hat{P}_1 )  |S ^ 0  (k) \rangle 
          + (k\to -k) 
          ,
\end{eqnarray} 
where $ |S ^ \mu \rangle $ has been defined in Eqs.~(\ref{si}),
(\ref{s0}) above.
Finally, using again Eqs.~\mref{trick1}, \mref{trick2} one finds the $ \widetilde{W}  $
propagator as 
\begin{eqnarray}
        \widetilde{W} 
        \,\, \epsfig{file=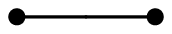,width=1.5cm}\,\, 
        \widetilde{W} 
        =
        \frac{ T}{\mmdebye }
        \langle \vec v  | 
        \left(1 - \hat{P}_0 -\hat{P}_1 \right)
        \hat{S}(k) 
        \left(1 - \hat{P}_0 - \hat{P}_1 \right)
        |\vec v ' \rangle
        + (k \to -k)
        \label{ww}
        .
\end{eqnarray}  

\section{Large momentum behavior of the propagators,  1$-$loop 
ultraviolet
  divergences, and renormali\-zability}
\setcounter{equation}0 
\label{sc:large} 


\begin{figure}[t]
\label{fg:divergent} 
\begin{center}
\begin{picture}(80,80)(0,0)

\hspace{-1cm}
\epsfig{file=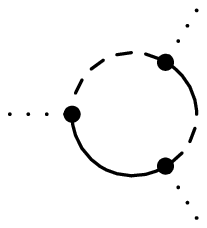,width=3cm}\hspace{2cm}

\end{picture}
\begin{picture}(80,80)(0,-12)

\hspace{1cm}
\epsfig{file=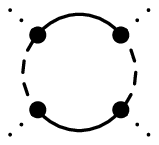,width=2.5cm}

\end{picture}
\end{center}
\caption[a]{Examples for
1-loop diagrams which are ultraviolet divergent,
and for which no counterterms are available in the action (\ref{action0}).
}
\end{figure}


In this section we first study the large momentum behavior of the 
propagators to see what kind of ultraviolet divergences can appear. 
The large momentum behavior of the $ A ^ 0 $ propagator, Eq.~(\ref{tree_a0a0}),
can be easily estimated. We have $ G _ \ell \sim 
1/k $, so that 
\begin{eqnarray}
  \langle A ^ 0 A ^ 0 \rangle \sim \frac{ 1}{k} 
        .
\end{eqnarray} 
The $  \widetilde{W} $ propagator (\ref{ww})  shows
the same behavior,
\begin{eqnarray}
  \langle  \widetilde{W} \widetilde{W}\rangle \sim \frac{ 1}{k} 
        .
\end{eqnarray} 
Now consider the transverse gauge field propagator (\ref{tree_aa})
\begin{eqnarray}
  \langle \vec A \vec A \rangle \sim \frac{ 1}{k ^ 0}
  \left[ \Delta (k) - \Delta (-k) \right] 
        .
\end{eqnarray} 
One can write the square bracket as 
\begin{eqnarray}
  \Delta (k) - \Delta (-k) = \frac{ 2 i m^2_{\rm D} k ^ 0 \re G\tr}
  {\left[ \vec k ^ 2 + k ^ 0 (m^2_{\rm D} - d \vec k ^ 2 )\im G \tr \right] ^ 2
    + k _ 0 ^ 2 (m^2_{\rm D} - d \vec k ^ 2 ) \re G\tr ^ 2 }
  \label{blT3} 
  .
\end{eqnarray} 
For  $ \vec k ^ 2 \gg m^2_{\rm D} $ one can simplify the denominator,
\begin{eqnarray}
  \Delta (k) - \Delta (-k) \simeq 
  \frac{ 1}{\vec k ^ 4}
  \frac{ 2 i m^2_{\rm D} k ^ 0 \re G\tr}
  {\left[ 1 - d  k ^ 0 \im G \tr \right] ^ 2
    + d ^ 2 k _ 0 ^ 2  \re G\tr ^ 2 }
  \label{blT4} 
        .
\end{eqnarray} 
For large momenta the products $ k ^ 0 \re G\tr $ and 
$ k ^ 0 \im G \tr $ are at
most of order unity. Thus for $ \vec k ^ 2 \gg m^2_{\rm D} $ 
 the transverse gauge field
propagator (\ref{tree_aa}) scales like
\begin{eqnarray}
  \langle \vec A \vec A \rangle  \sim \frac{1}{ k ^ 0 \vec k ^ 4}
        \label{falloff} 
        .
\end{eqnarray} 
It is a little puzzling that this asymptotic behavior sets in 
only when $ \vec k ^
2\gg \mmdebye $, which is well outside the range of validity 
of Eq.~(\ref{boltzmann}), $ \vec k ^ 2 \ll m^2_{\rm D} $. 

A consequence of Eq.~(\ref{falloff}) is that the diagrams 
(\ref{d1}), (\ref{d2}),
(\ref{d4})  in Appendix~\ref{ap:diagrams}, which contain transverse
gauge field propagators, would be ultraviolet finite when computed with
the propagator  (\ref{tree_aa}). The results in Appendix~\ref{ap:diagrams} 
are obtained by using  the approximated propagators of Appendix
\ref{ap:estimates} giving ultraviolet divergent results (see the
discussion at the end of Sec.~\ref{sc:nllo}). 

\subsection{Renormalizability}
\label{sc:renormalizability} 

Now we address the question whether Eq.~(\ref{boltzmann}) is a 
renormalizable effective theory. As we have seen above, 
 certain propagators, like $ \langle
\widetilde{W} \widetilde{W} \rangle $ and $ \langle A ^ 0 A ^ 0 \rangle $, fall
off only like $ 1/k $ for $ k \to \infty $. Therefore diagrams with 3 and 4
external $ \lambda $ lines,  which contain only these propagators (see 
Fig.~\ref{fg:divergent}), 
 are linearly and logarithmically divergent.  The terms in the
action (\ref{action0}), on the other hand, 
are at most quadratic in $ \lambda $.
Thus in order to renormalize these divergences one would have to introduce new
terms in the action. But then the path integral (\ref{path}) 
 would  no longer be equivalent to the 
Langevin equation  (\ref{boltzmann}). 
Thus
it would not be possible to use the theory (\ref{boltzmann}) for lattice
simulations if one wants to take the continuum limit
\footnote{
It is of course possible that these
divergent contributions cancel among different diagrams. However, I do 
not see a
reason why such a cancellation should occur. 
}.
If one is interested
in calculating corrections to the
leading log Langevin equation in an expansion in $ [\log(1/g)] ^{-1} $
this does not pose a problem since the
divergent terms would contribute at higher orders in $ g $. 

\section{The logarithmic approximation beyond leading order}
\setcounter{equation}0 
\label{sc:nllo} 

Arnold and Yaffe have argued that the leading log Langevin equation
(\ref{langevin}) is also valid at next-to-leading logarithmic order
(NLLO). 
We will now present a simple argument why this should indeed be the case. 

Let us first recall the argument which takes us from Eq.~(\ref{boltzmann}) to
Eq.~(\ref{langevin}) in the {\em leading} log approximation.

For the
non-perturbative fields with spatial momenta of order $ g ^ 2 T $ the collision
term, which is of order $ g ^ 2 \log(1/g) T $, is large compared to the term
containing $ \vec v \cdot \vec D $. If one neglects the latter, the $ l = 1 $
projection of Eq.~(\ref{boltzmann}) does not contain $ \widetilde{W} $ and
is thus a closed equation for the gauge fields alone 
\footnote{For a detailed
discussion  see \cite{effective}.}, which is 
precisely Eq.~(\ref{langevin}). 
Similarly, the equations for
different $ l $ and $ m $ components of $ \widetilde{W} $ decouple in the 
leading log approximation. 

Now imagine continuing this approximation scheme by expanding in powers of $
L ^{-1} \equiv \log(1/g) ^{-1}  $. For the moment we ignore  the 
Fourier components with $ \vec k \sim L g ^ 2  T$. At NLL
order the $ l = 1 $ projection of Eq.~(\ref{boltzmann}) contains a term which
schematically looks like $ \vec v \cdot \vec D \widetilde{W} _{l=2} $. At this
order we can use the leading log approximation for $ \widetilde{W} _{l=2} $ to
write 
\begin{eqnarray}  
  \vec v \cdot \vec D \widetilde{W} _ {l=2} \sim \vec v \cdot \vec D C ^
  {-1} \xi _{l=2} \sim \log(1/g)  ^{-1} \xi _{l=2} 
  \label{correction} 
  .
\end{eqnarray} 
Thus it would appear that the correction term is suppressed by only a single
power of $ L ^{-1} $. However, since the correlation functions
of $ \xi  $ are diagonal in the $ (l,m) $ basis (see Eq.~(\ref{xicorrelator})),
one needs at least two factors of $  \xi _{l=2} $ to obtain a non-vanishing
contribution. But then one also gets two factors of $  L ^{-1} $, 
which shows that the correction term in Eq.~(\ref{correction}) would become
relevant only at next-to-next-to-leading log order. 
Thus even at NLLO  we can ignore the  
the $ l \ge 2 $ projections of
Eq.~(\ref{action0}), and we can drop $ \widetilde{W} $ field 
from 
(\ref{action0}). Then we obtain the action
\begin{eqnarray}
  S_{\rm b} = \int d ^ 4 x 
  \left\{ \frac{ T}{\gamma_{\rm b} } \gvec{\lambda  } ^ 2
    - i \gvec{\lambda  }
    \cdot \left( \frac{ 1}{\gamma_{\rm b} } \vec D \times  \vec B
      - \vec E \right) \right\} 
  \label{sbare} 
\end{eqnarray} 
with
\begin{eqnarray}
  \gamma_{\rm b} \equiv \frac{ m^2_{\rm D} }{d c _ 1}
  \label{gammabare} 
  .
\end{eqnarray}

The modes with spatial momenta of order of the collision term, 
$ \vec k  \sim g ^ 2 L T $, \footnote{For simplicity we will 
oftentimes
write $ \vec k \sim C $ instead of $ \vec k  \sim g ^ 2 L T $.}
enter the soft non-perturbative 
dynamics through loops.
They can 
be treated perturbatively, where the loop expansion parameter is $
L ^{-1} $. They change the coefficients
in the action which becomes
\begin{eqnarray}
  S = \int d ^ 4 x \left\{ \frac{ T}{\gamma_{\rm b} } 
    Z _ \lambda  \gvec{\lambda  } ^ 2
    - i \gvec{\lambda  }
    \cdot \left( \frac{ 1}{\gamma_{\rm b} } Z _ B \vec D \times  \vec B
      - Z _ E \vec E \right) \right\} 
  \label{sren} 
  .
\end{eqnarray}
The $ k \sim C $ modes will also introduce higher dimensional
operators in Eq.~(\ref{sren}). These, however, are suppressed by powers of
$ g ^ 2 T/C \sim L ^{-1} $.

The Lagrange multiplier field $ \gvec \lambda $ can be  
rescaled to bring the action back into the form (\ref{sbare}), 
\begin{eqnarray}
  \gvec{\lambda  } \to Z_E ^{-1}  \gvec{\lambda  }
  \label{blR2} 
  ,
\end{eqnarray} 
so that Eq.~(\ref{sren}) becomes
\begin{eqnarray}
  S = \int d ^ 4 x \left\{ \frac{ T}{\gamma_{\rm b} } 
    \frac{ Z _ \lambda }{Z _ E ^ 2}  \gvec{\lambda  } ^ 2
    - i \gvec{\lambda  }
    \cdot \left( \frac{ 1}{\gamma_{\rm b} } \frac{ Z _ B}{Z _ E}
      \vec D \times  \vec B
      -  \vec E \right) \right\} 
  \label{sren2} 
  .
\end{eqnarray}
Now we can read off the
color conductivity 
$ \gamma $ which takes into account of the $ \vec k \sim g ^
2 L T $ modes  from the coefficient of the second term in
Eq.~(\ref{sren}),
\begin{eqnarray}
  \gamma  = \frac{ Z _ E}{Z _ B} \gamma_{\rm b}
  .
\end{eqnarray}
Alternatively one could read off $ \gamma $ from the first term in
Eq.~(\ref{sren}) 
because we already know that 
both Eq.~(\ref{sren}) 
and Eq.~(\ref{boltzmann})  generate a thermal 
ensemble of magnetic scale gauge fields at temperature $ T $ 
\cite{zj,damgaard,local} (see also
Eq.~(\ref{hamiltonian})). 
In other words, the temperature $ T $ in Eq.~(\ref{sren2}) is the same
temperature appearing in Eq.~(\ref{boltzmann}). Therefore we must have 
\begin{eqnarray}
  \gamma  = \frac{ Z _ E ^ 2}{Z _ \lambda } \gamma_{\rm b}
  .
\end{eqnarray} 
Thus  the $ Z $-factors must be related by
\begin{eqnarray}
  Z _ \lambda  = Z _ E Z _ B
  ,
\end{eqnarray} 
and  the action 
contains only one unknown  physical parameter $ \gamma  $. Thus in order to
determine  $ \gamma  $ rather than to calculate the $ Z $-factors
 it is sufficient to calculate 
one single quantity  in the theories (\ref{boltzmann}) and
(\ref{langevin}) and require that the results match.  

We have seen in Sec.~\ref{sc:large} that loops containing gauge
field propagators in general receive non-trivial contributions from
momenta of order $ m_{\rm D} \sim g T $. These will not be included
in the calculation of $ \gamma  $ in Sec.~\ref{sc:wilson}, i.e., the
calculation in Sec.~\ref{sc:wilson}  is performed with an ultraviolet
cutoff separating the scales $ g T $ and $ g ^ 2 T $. Technically this
is achieved in dimensional regularization by not using the propagators of
Sec.~\ref{sc:pathbl} but the approximated expressions for $ \vec k ^ 2 
\ll m^2_{\rm D} $. These are listed in Appendix \ref{ap:estimates}.

\section{Determining $ \gamma  $ by calculating Wilson loops}
\setcounter{equation}0 
\label{sc:wilson} 

To determine the next-to-leading log color conductivity $ \gamma  $
we follow Arnold and Yaffe \cite{ay:nllo} and match
the results for a rectangular Wilson loop $ {\cal W }(t,R) $, with 
spatial
extent $ R $ and temporal extent $ t $,  
in the theory (\ref{sren})  and in the underlying "microscopic" theory, 
which in our case is Eq.~(\ref{boltzmann}). 
The Wilson loop is defined as 
\begin{eqnarray}
  {\cal W }(t,R) = d_{\rm R}^{-1} 
  \mbox{tr} \left \langle \mbox{P} \exp \left\{ 
      i g \int d x \cdot A \right\} \right \rangle 
  ,
\end{eqnarray}
where P denotes path ordering, and the 
integration is along a closed contour starting at $ x ^ \mu  
= 0 $, then going
a  distance $ R $ in the $ z $- direction, then a distance $ t $ in
the time direction, back in the $ z $-direction to $ z=0 $ and finally
back to $ x ^ \mu  = 0 $ along the time direction. 
The normalization factor $d_{\rm R}^{-1}$, where $ d_{\rm R} $ is the 
dimension of the representation R associated with the Wilson loop, is 
included so that $ |{\cal W }| \leq   1 $.

One has to choose $ R $ such that $ R ^{-1}  \gg g ^ 2 T $, so that
one can obtain perturbative contributions to $ {\cal W } $ 
(see Sec.~\ref{sc:w2ll}). Furthermore, one needs $ R ^{-1} \ll
g ^ 2 \log(1/g) T $, where both theories (\ref{boltzmann}) and (\ref{langevin}) 
are valid. 
Unlike in \cite{ay:nllo} we do not consider the limit $ t \to \infty $. 
We first consider Coulomb gauge and later the general flow gauge
(\ref{flowgauge}).   


At lowest non-trivial order
the
contributions to $  {\cal W } $ are due to propagators connecting the loop
with itself. As in \cite{ay:nllo}  we ignore contributions connecting
one edge with itself, these depend either only on $ t $ or 
only on $ R $. In Coulomb gauge there are no diagrams which connect a
space-like edge with a time-like one. Thus at second order in the fields
there are  two contributions, 
\begin{eqnarray}
  {\cal W }^{(2)}  (t, R) = 
  {\cal W }^{(2)} _{\rm time} (t, R) + {\cal W }^{(2)} _{\rm space} (t, R)
  ,
\end{eqnarray}  
with 
\footnote{In Ref.~\cite{ay:nllo} there is a sign error  in the
analogue of Eq.~(\ref{w2time}). Exponentiating Eq.~(\ref{w2time}) (with the
correct sign) one would find  that  $ {\cal W } $ {\em grows} exponentially
for $ t \to \infty $. 
This error does, however, not affect the result
for the color conductivity because the wrong sign it is used consistently. }
\begin{eqnarray} 
  {\cal W }^{(2)} _{\rm time} (t, R) &=&  
  g ^ 2 C_{\rm R} \int _ 0 ^ t d t _ 1 
  \int _ 0 ^ t d t _ 2 \langle A ^  0  (t _ 1, \vec
  0) A ^  0  (t _ 2, R \vec {\hat{z}  }) \rangle
  \label{w2time}  
  ,
\\
  {\cal W }^{(2)} _{\rm space} (t, R) &=& 
  g ^ 2 C_{\rm R} 
  \int _ 0 ^ R d z _ 1 \int _ 0 ^ R d z _ 2 \langle A ^ z (0, 
  z _ 1 \vec{\hat{z}}  ) A ^ z (t , z _ 2 \vec {\hat{z} }) \rangle 
  ,
\end{eqnarray} 
where $ C_{\rm R} $ is the quadratic
Casimir operator of the representation R. 

\subsection{$ {\cal W }^{(2)} $ in the leading log Langevin equation}
\label{sc:w2ll} 

Calculating $ {\cal W } (t, R) $ from in the theory (\ref{langevin}) 
is straightforward. 
The propagators are the ones for Eq.~(\ref{boltzmann}) in the limit
that  $ k ^ 0 \sim g ^ 4 T $ (modulo logarithms) and  $ |\vec k|\ll C $.
The relevant expressions  are listed in Appendix \ref{sc:k0g4t}. 
Here $ \gamma  $ must be understood as the 
complete NLLO color conductivity which includes the effect of the
modes with $ \vec k \sim C $. Thus we have
\begin{eqnarray} 
        A ^ 0
        \,\, \epsfig{file=a0a0prop.eps,width=1.5cm}\,\, 
        A ^ 0
        &=& 
        \frac{ 2T}{\gamma  \vec p ^ 2}    
        \label{treea0ll} 
        ,
\\
        A ^ i
        \,\, \epsfig{file=aaprop.eps,width=1.5cm}\,\, 
        A ^ j    &=& 
        T P\tr ^{ij}
         \frac{- i}{p ^ 0}  \frac{ 1}{\vec p ^ 2 - i \gamma  p ^ 0}
         + (p \to -p)
        \label{treeaall}
        .
\end{eqnarray}
First consider $ {\cal W }^{(2)} _{\rm time} $.
The integral over $ p $ is saturated by momenta $ |\vec p | \sim R ^{-1}  $,
and  one obtains
\begin{eqnarray} 
  {\cal W }^{(2)} _{\rm time} (t, R) = \frac{ C_{\rm R} g ^ 2 T }
  {2 \pi   } \frac{ t}{\gamma R}
  \label{blWL3.1} 
  .
\end{eqnarray}
To compute $ {\cal W }^{(2)} _{\rm space} $ we first perform the
integration over $ z _ 1 $ and $ z _ 2 $, then the $ p ^ 0 $-integration
which gives
\begin{eqnarray} 
  {\cal W }^{(2)} _{\rm space}(t, R) = - \frac{ C_{\rm R} g ^ 2 T }{2\pi  ^ 2}
  \int _ 0 ^ \infty \frac{ d p }{p ^ 2} e ^{-p ^ 2 t/\gamma  } 
  \int _ {-1} ^ 1 d \cos\theta   \frac{ 1 - \cos ^ 2 \theta  }
  {\cos ^ 2 \theta }\left[ 1 - \cos(p R \cos\theta ) \right]  
  \label{WL31.1} 
  .
\end{eqnarray} 
This integral can be simplified if we choose
\begin{eqnarray} 
  \frac{ t}{\gamma }  \gg R ^ 2 
  \label{larget} 
  .
\end{eqnarray} 
Then the dominant
contribution is from the integration region
$ p ^ 2 \sim \gamma  /t \ll R ^ {-2} $.
For Eq.~(\ref{WL31.1}) to be reliable 
we need $ p \gg g ^ 2 T $, i.e., 
$ t/\gamma  \ll (g ^ 2 T )^ {-2 }$. Since $ p R \ll 1 $ the 
square bracket can be expanded around $ p R = 0 $, and to  lowest non-trivial
order in this expansion we obtain
\begin{eqnarray}
   {\cal W }^{(2)} _{\rm space} = - \frac{ C_{\rm R} g ^ 2 T}{6 \sqrt{\pi  }}
   \frac{ t}{\gamma  R} \left( \frac{ \gamma  R ^ 2}{t} \right) ^ \frac{ 3}{2}
   .
\end{eqnarray}
Due to Eq.~(\ref{larget})  this 
is small compared to the result for the temporal contribution 
(\ref{blWL3.1}). 

Higher loops will receive contributions from momenta of order $ g ^ 2 T $, and
are thus not calculable in perturbation theory. By construction, these
are the same in both theories (\ref{boltzmann}) and (\ref{langevin}) and
can be discarded for a matching calculation. This is achieved by using
dimensional  to cut off both ultraviolet and infrared divergences. 
With this regularization loop corrections to Eqs.~(\ref{blWL3.1}) vanish.
Then  the leading non-trivial contribution to  $ {\cal W }^{(2)} $ is given by 
Eq.~(\ref{blWL3.1}) provided that
\begin{eqnarray} 
  g ^ 2 T \ll \frac{ \gamma  }{t} \ll R ^ {-1} \ll C 
  .
\end{eqnarray}

\subsection{$ {\cal W }^{(2)} $ in the Boltzmann-Langevin equation}
\label{sc:w2bl}

Now we compute
$  {\cal W }^{(2)} $ in the theory (\ref{boltzmann}) and 
match the result with Eq.~(\ref{blWL3.1}) to determine
the next-to-leading log color conductivity $ \gamma  $. We have
seen in Sec.~\ref{sc:w2ll} that in Coulomb gauge $  {\cal W }^{(2)} $ 
is determined by the $ A ^ 0 $-propagator.
The lowest order contribution is again given Eq.~(\ref{blWL3.1}) 
but with $ \gamma  $ replaced by the 'bare' value $ \gamma  _{\rm b} $.
Now loop corrections to this result do not vanish, and we have to compute
the 1-loop contributions to the $ A ^ 0 $-propagator.

\begin{figure}[t]


\hspace{.5cm}

\raisebox{-.3cm}{\epsfig{file=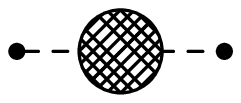,width=2cm}}
=
\raisebox{0cm}{\epsfig{file=a0a0prop.eps,width=1.5cm}}
+
\raisebox{-.35cm}{\epsfig{file=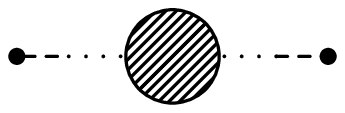,width=3cm}}
+
\raisebox{-.35cm}{\epsfig{file=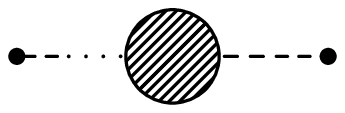,width=3cm}}
+
\raisebox{-.35cm}{\epsfig{file=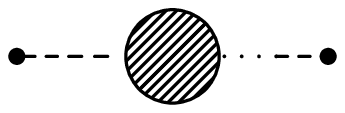,width=3cm}}

  \caption{The full $ A ^ 0$ propagator in Coulomb gauge.}
\label{fg:a0propagator}   
\end{figure}

First let us note that the $ A A $ self-energy is zero.  Every
interaction vertex involves one $ \lambda $ field. Thus each propagator in the
loop would be a propagator which mixes $ \lambda $ with a physical field.  Then
all propagators in the loop have singularities only in the lower half of the
complex $ k ^ 0 $ plane
when $ k $ is the loop momentum.  
Consequently, the $ k ^ 0 $-integration contour can be moved to
$ i \infty $, and the integral vanishes. One can also understand this as a
result of the invariance of the effective action $ \Gamma  $, 
the
generating functional of 1-particle irreducible Green functions, 
under the BRST-transformation (\ref{brs}). 
$ \Gamma  $ is invariant because the because the
transformation (\ref{brs}) is linear. 
A term $ \half \varphi ^ i \Pi ^{ij}
\varphi ^{j} $ in  $ \Gamma  $ 
would transform into $ \varphi ^ i \Pi
^{ij} \bar{\epsilon }\eta ^ j$, and there is no 
other term which transforms into
this, so it must vanish.

Thus the full $ A ^ 0 $-propagator only receives contributions from
the $ \lambda \lambda $- and $ A ^ 0 \lambda $-selfenergies (see 
Fig.~\ref{fg:a0propagator}).  For small momenta
$ A ^ 0 $ decouples from the $ l \ge 2 $ components of $ \lambda $ (see
Eq.~(\ref{s0th3})), and we only
have to consider the selfenergies for the $ l = 1 $ or vector 
part $ \lambda ^ i
= \langle ^ i|\lambda \rangle $.  The diagrams for $ \Pi _{\lambda \lambda } $
are depicted in Appendix \ref{ap:diagrams}.  For loop momenta with $ k ^ 0 \sim
C ^ 3 /m^2_{\rm D}$, $ \vec k \sim C $ they contribute
\begin{eqnarray}
  \Pi  _{
    \lambda  \lambda  } \sim g ^ 2 T C ^ {-1} \times \frac{ T C}{m^2_{\rm D} }
  \sim \log(1/g)  ^{-1} \times \frac{ T C}{m^2_{\rm D} }
  \label{blN7.2.0.2a} 
  .
\end{eqnarray}  
This has to be compared with 
the coefficient of the $ \lambda  \lambda  $-term in the 
action (\ref{action}) which is of order $
TC/m^2_{\rm D} $. Therefore $ \Pi  _{
    \lambda  \lambda  } $ yields a contribution to the $ A ^ 0 $-propagator
which is of next-to-leading order in $  \log(1/g)  ^{-1} $. 
In addition to the diagrams in  Appendix \ref{ap:diagrams} there are diagrams
involving the $ \lambda A ^ 0
\widetilde{W} $ vertex. 
Using the estimates (\ref{estimates1}), (\ref{estimates2}) one easily
sees that these are suppressed compared to Eq.~(\ref{blN7.2.0.2a})
by  powers of $ g $.

The contributions due to loop momenta $ k ^ 0 \sim \vec k \sim C $ are easily
estimated since then all propagators are of the same order of magnitude (see
Eq.~(\ref{estimates1})). One finds
\begin{eqnarray} 
    \Pi  _{
    \lambda  \lambda  }\Big|  _{{\rm loop \, frequency} \sim C 
  } 
  \sim \frac{ g ^ 2 T C}{m^2_{\rm D}} \times \frac{ TC}{m^2_{\rm D} } 
  \sim g ^ 2  \log(1/g)   \times \frac{ T C}{m^2_{\rm D} }
  \label{blN7.2.0.2b} 
  ,
\end{eqnarray}  
which is suppressed compared to   (\ref{blN7.2.0.2a}) by two powers
of $ g $. Thus, unlike for the $ \vec A \vec A $-selfenergy
to be 
discussed in Sec.~\ref{sc:spatial}, the loop frequencies of order  $ C $
can be neglected here. 

Now consider the $ \lambda  A ^ 0 $-selfenergy. For the diagrams listed in
Appendix \ref{ap:diagrams}   we estimate $ \Pi  _{\lambda  A ^ 0}  \sim g ^ 2 T
  $ for loop momenta with $ k ^ 0 \sim C ^ 3/m^2_{\rm D} $. For loop momenta
  with
$ k ^ 0 \sim C $ we find  $ \Pi  _{\lambda  A ^ 0}  \sim g ^ 4 T
  $ modulo $ \log(1/g) $. 
Again the contributions which contain the $ \lambda  A ^ 0
\widetilde{W} $-vertex are suppressed. 

Thus we can use the simplified form of the propagators from Sec.~\ref{sc:k0g4t}
to compute the loop integrals.  For the external lines we can use the
approximations in Sec.~\ref{sc:k0g4tl}.  In Coulomb gauge we find that the 
$ A ^ 0 $ propagator for $ p ^ 0 \ll C ^ 3/m^2_{\rm D} $ and
$ |\vec p | \ll C $  is given by 
\begin{eqnarray} 
\raisebox{-.3cm}{\epsfig{file=a0propagator1.eps,width=2cm}}
  &=& 
\raisebox{0cm}{\epsfig{file=a0a0prop.eps,width=1.5cm}}
+
  -\frac{ p ^ i p ^ j}{\vec p ^ 4} \times 
  \raisebox{-.3cm}{\epsfig{file=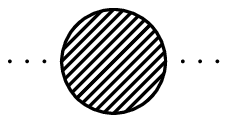,width=1.5cm}}
        + 4 d \frac{ T c _ 1}{m^2_{\rm D}} \frac{ p ^ i }{\vec p ^ 4}
        \times 
  \raisebox{-.3cm}{\epsfig{file=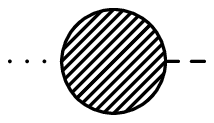,width=1.5cm}}
  \nonumber \\ &=& 
\frac{ 2 T }{\gamma_{\rm b} \vec p ^ 2 }+
  \frac{ 1}{\vec p ^ 2} \Pi _{{\lambda}{\lambda}} (0)
  - \frac{ 4 d c _ 1 T }{m^2_{\rm D} } \frac{ p ^ i }{\vec p ^ 4}
  \left. \frac{ \partial \Pi _{{\lambda} A ^ 0} ^ i }{\partial p ^ j}
  \right| _{p = 0}
  p ^ j
  \label{blR8}
  \\ &=& 
\frac{ 2 T }{\gamma_{\rm b} \vec p ^ 2 }+
  2 d \frac{ N g ^ 2 T ^ 2}{m^2_{\rm D} \vec p ^ 2} 
  \int _{\vec k } \Bigg\{ \frac{ 1}{\vec k ^ 2}
          P\tr ^{lm}   (\vec k)    \langle ^0 | \hat{v} ^ k \hat{v} ^ l 
  \hat{G} (0,\vec k) \hat{v} ^ k \hat{v} ^ m |^0 \rangle 
  \nonumber \\ &&{}
   +(d-1) c _ 1 ^ 2 \frac{ G\tr (0, \vec k)}
  {\vec k ^ 4} + 4 \frac{ d -1}{d} \frac{ c _ 1}{\vec k ^ 4}
  \left[ 1 - d c _ 1  G\tr (0, \vec k) \right] \Bigg\} 
  \nonumber 
  .
\end{eqnarray}
Comparing this with the corresponding expression for the NLLO Langevin
equation (\ref{treea0ll}) we find 
\begin{eqnarray}
  \frac{ 1}{\gamma  } &=&  \frac{ 1}{\gamma_{\rm b} } +
   \frac{d N g ^ 2 T }{m^2_{\rm D}} 
  \int _{\vec k } \Bigg\{ \frac{ 1}{\vec k ^ 2}
          P\tr ^{lm}   (\vec k)    \langle ^0 | \hat{v} ^ k \hat{v} ^ l 
  \hat{G} (0,\vec k) \hat{v} ^ k \hat{v} ^ m |^0 \rangle 
  \nonumber \\ &&{}
   +(d-1) c _ 1 ^ 2 \frac{ G\tr (0, \vec k)}
  {\vec k ^ 4} + 4 \frac{ d -1}{d} \frac{ c _ 1}{\vec k ^ 4}
  \left[ 1 - d c _ 1  G\tr (0, \vec k) \right] \Bigg\} 
  \label{R20} 
  .
\end{eqnarray}
for the NLLO color conductivity $ \gamma  $ 
in agreement with the result of Arnold and Yaffe \cite{ay:nllo} \footnote{
Arnold and Yaffe \cite{ay:nllo}  write their results in terms of 
$ \sigma _ {\vec k} = \bar{\sigma} _{\rm T}^{(0)}(\vec k)$, which
is related to our notation by 
$ 
  \sigma _ {\vec k} = \bar{\sigma} _{\rm T}^{(0)}(\vec k) 
  = m^2_{\rm D} G\tr(0,\vec k)
$. Furthermore, their $ \gamma  _ 1 $ is the same as our $ c _ 1 $.}. 
We have not checked the numerical evaluation of $ \gamma  $ 
performed
in \cite{ay:nllo}.

\begin{figure}[t]
 
  \begin{center}
 \raisebox{-.35cm}{\epsfig{file=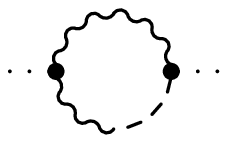,width=3cm}}
\hspace{1cm}
\raisebox{-.35cm}{\epsfig{file=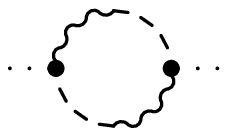,width=3cm}}
\hspace{1cm}
\raisebox{-.35cm}{\epsfig{file=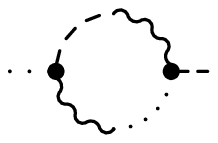,width=3cm}}
  \end{center}

  \caption{Examples for additional contributions to the  $ \Pi  _{\lambda
      \lambda  } $  and $ \Pi  _{\lambda  A ^ 0  } $ self-energies
in the flow gauge (\ref{flowgauge})  with $ \kappa  \neq 0 $.}
  \label{fg:blU1} 
\end{figure}

\subsection{Gauge fixing dependence}
\label{sc:gf} 

In this section we address the question whether the above result for the Wilson
loop and thus for the color conductivity depends on our gauge choice
for the theory (\ref{boltzmann}).  
Consider the gauge (\ref{flowgauge}) for $ \kappa  \neq 0 $. 
One could leave the parametric size of $ \kappa  $ arbitrary, 
but for the purpose of power counting
it is a lot more convenient to adopt a natural choice in terms of some 
scale which is already present in Eq.~(\ref{boltzmann}). 

In the leading log theory
(\ref{langevin}) there would be only one natural choice. The characteristic
frequency of the transverse gauge fields is $ k ^ 0 \sim \vec k ^ 2/\gamma $,
which can be seen from Eq.~(\ref{treeaall}).
In Coulomb gauge the  $ A ^0 $-propagator, 
Eq.~(\ref{treea0ll}),  is frequency-independent.
By choosing $ \kappa 
\sim \gamma  $ one would make the characteristic frequencies of all gauge
field components equal at tree level. 

For Eq.~(\ref{boltzmann}) there is an additional option.
Consider the $ A ^ 0 $ propagator in 
Eq.~(\ref{tree_a0a0}). 
$ G _ \ell (k)$  has discontinuities for $ k ^ 0 $ of order of the
collision term $ C $, and so does the $ A ^ 0 $-propagator. 
When $ \vec k \sim C $, the characteristic frequency of the transverse
gauge fields is of order $ C $ as well. Thus  one
would naturally use $  \kappa  \sim C $. This case turns out to be very 
easy to analyze. 
As long as $ k ^ 0 \sim g ^ 4 T $ (modulo logs) 
all propagators involving $ A ^ 0 $ are approximately the
same as in Coulomb gauge, and we do not have to recalculate any
of the diagrams in Appendix \ref{ap:diagrams}. 
Furthermore, even for $ k ^ 0 \sim C $,  $ A ^ 0 $ is of the same order
of magnitude as in Coulomb gauge. Therefore the suppression of the
integration region $ k ^ 0 \sim C $ found in Sec.~\ref{sc:w2bl}  persists. 
There are additional diagrams contributing to the
$ \Pi  _{\lambda  \lambda  } $ and $ \Pi  _{\lambda  A ^ 0 }  $ self-energies
due to the mixing of $ A ^ 0 $ and $ \vec A $, see Fig.~\ref{fg:blU1}. 
Using the estimates of Appendix \ref{ap:estimates} one easily finds
that all these contributions are suppressed by powers of $ g $. 
There are also contributions to the 1-loop $ A ^ 0 $-propagator of the
form
\begin{eqnarray}
        \langle A ^ 0 A ^ 0 \rangle _{\rm 1-loop}
        \sim 
        \langle A ^ 0 \vec A _ \ell \rangle    
         \Pi _ {\vec A \lambda } \langle \lambda A ^ 0  \rangle 
         \label{mixing} 
         .
\end{eqnarray}
The longitudinal gauge field $ \vec A _\ell $ is of the same order of magnitude
as $ A ^ 0 _{\rm Coul} $ when $ \vec k \sim C $, and of order $ \kappa/|\vec k|
\sim C/(g ^ 2 T) $ times $ A ^ 0 _{\rm Coul} $ when $ \vec k \sim g ^ 2 T $.
Therefore $ \langle A ^ 0 \vec A _ \ell \rangle \lsim \langle A ^ 0 A ^ 0
\rangle $.  The self-energy $ \Pi _ {\vec A \lambda }$ is estimated in
Sec.~\ref{sc:spatial} as $ (g ^ 2 T) ^ 3/m^2_{\rm D} $ modulo logs. Finally,
from Eq.~(\ref{estimates3})   $ \langle \lambda A ^ 0  \rangle \sim 
(g ^ 2 T )^{-1} $, so that
\begin{eqnarray}
        \langle A ^ 0 \vec A \rangle    
         \Pi _ {\vec A \lambda } \langle \lambda A ^ 0  \rangle 
        \lsim 
        \frac{ ( g ^ 2 T)^ 2}{m^2_{\rm D}}\langle A ^ 0 A ^ 0 \rangle
\end{eqnarray}
which is strongly suppressed compared to the tree level result for 
$ \langle A ^ 0 A ^ 0 \rangle $. 
Thus we conclude that the result for the Wilson loop in the 
theory (\ref{boltzmann})  is
not changed if one uses the flow gauge with $ \kappa  \sim C $ instead
of Coulomb gauge \footnote{Things become more complicated for
$ \kappa  \sim \gamma  $. In this case Eq.~(\ref{mixing}) would contribute
to the NLLO result for the Wilson loop. Furthermore, one would have to 
take into account not only self-energy corrections to the $ A ^ 0 $-propagator
but also vertex corrections.}.

\section{Transverse gauge field propagator at one loop}
\setcounter{equation}0 
\label{sc:spatial} 

We have seen that the NLLO color conductivity can be obtained by considering
only the $ A ^ 0 $-propagator which gives the dominant contribution to
the Wilson loop of Sec.~\ref{sc:wilson}. By simple power counting we have
found that the contribution from loop momenta $ k $ with
\footnote{In this section we ignore $ \log(1/g) $ in the order of magnitude
estimates, so that, e.g., the collision term $C$ is estimated as
$ C \sim g ^ 2 T $.}  
$ k ^ 0 \sim g ^ 2 T $ is suppressed by {\em powers} of $ g $. 
Thus it appears that the dynamics associated with this frequency scale
(see Ref.~\cite{local}) is irrelevant to the soft gauge field dynamics.  
From 
the discussion in Sec.~\ref{sc:nllo}  it should be clear that one
could have chosen another quantity for the matching calculation which
receives contributions from spatial gauge fields as well. 

In this
section we will try to see whether the spatial gauge field propagator
is affected by loop momenta with $k ^ 0 \sim g ^ 2 T $,
when external momenta are of order 
$ \vec p \sim g ^ 2 T $, $ p ^ 0 \sim g ^ 4 T $. 
It cannot be calculated perturbatively, but
one can calculate the contribution from loop momenta large compared
to $ g ^ 2 T $, 
which is the   philosophy of Ref.~\cite{ladder}.


\begin{figure}[t]
\label{fg:transverse} 
\begin{center}
\begin{picture}(80,80)(0,0)

\hspace{-3cm}
\epsfig{file=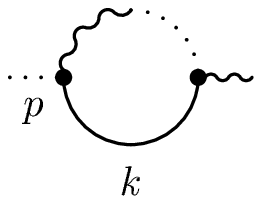,width=3cm}\hspace{2cm}
\epsfig{file=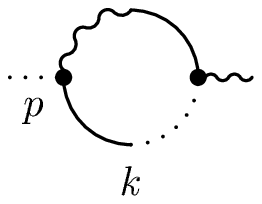,width=3cm}

\end{picture}
\end{center}
\caption[a]{One loop contributions to the $ \vec A $-$ \lambda $ 
self-energy which receive equal contributions from 
loop momenta with $ k ^ 0 $ 
of order $ g ^ 2 T $ and of order $ g ^ 4 T $.}
\end{figure}


As we discussed in Sec.~\ref{sc:w2bl} the 
$ A A $ self-energy vanishes.  
Thus we only obtain contributions which contain the $ \lambda \vec A $ and
$ \lambda \lambda $ self-energies. The latter has already  been considered in
Sec.~\ref{sc:nllo}. Consider now the contribution of the  $ \lambda \vec A $ 
self-energy depicted in Fig.~\ref{fg:transverse} and use the estimates
in Eq.~(\ref{estimates2}). 
The $ \vec A \lambda $ propagator is of order $ C ^{-1} $ and
$ m^2_{\rm D} /C ^ 3 $ for $ k ^ 0 \sim C $ and 
$ k ^ 0 \sim  C ^ 3 /m^2_{\rm D} $, respectively. 
Therefore for both these frequency scales 
the product $ k ^ 0 \langle \vec A \lambda
\rangle $ is of the same order of magnitude $ k ^ 0 \langle \vec A \lambda
\rangle \sim 1 $. The 
other propagator in the loop is of the same order for both frequencies, 
$ \langle \widetilde{W} \widetilde{W} \rangle \sim T/(m^2_{\rm D} C) $. 
The integration over 3-momenta is of order $ C ^ 3 $, and there is
a factor $ g ^ 2 $ from the vertices. Thus we obtain $ \Pi _{\vec A \lambda } 
\sim  g ^ 2 T C ^ 2 /m^2_{\rm D} $. This would give a contribution
to the $ \vec A \vec A$ propagator of order (cf. Eq.~(\ref{estimates3})) 
\begin{eqnarray}
  \langle \vec A \vec A  \rangle _{\rm 1-loop} (p)
  \sim \langle \vec A \vec A \rangle \Pi _{\vec A \lambda }
  \langle  \lambda\vec A \rangle \sim
  \frac{ C}{ g ^ 2 T}\langle  \vec A  \vec A \rangle 
\end{eqnarray} 
for $ \vec p \sim g ^ 2 T $, $ p ^ 0 \sim C(g ^ 2 T)^ 2/m^2_{\rm D} $.
This would mean that the 1-loop contribution is larger than the
tree level by one power of $ \log(1/g) $. 
However, we would expect  that
 $   \Pi _{\vec A \lambda } (p) $
vanishes for $ p \to 0 $, and that one can expand
\begin{eqnarray}
  \Pi _{\vec A  \lambda }  (p) \sim 
  \vec p ^ 2  \Pi _{\vec A \lambda }' + 
  \cdots 
  \nonumber 
  .
\end{eqnarray}
Since we consider loop momenta $ k $ with $ \vec k \sim C $,
$ \Pi  ' $ is of order $ \Pi / C ^ 2 $. 
This would give the estimate
\begin{eqnarray}
  \langle \vec A\vec A  \rangle _{\rm 1-loop} (p)
  \sim \frac{ g ^ 2 T}{ C}\langle  \vec A \vec A \rangle
  .
\end{eqnarray} 
In other words, the diagram in Fig.~\ref{fg:transverse} would contribute at
next-to-leading logarithmic order.  
This is puzzling since 
the next-to-leading 
order color conductivity $ \gamma  $ which was calculated 
from the $ A _0 $ propagator alone does not receive such contributions.
It was argued in Sec.~\ref{sc:nllo}, based on the fact that 
Eq.~(\ref{boltzmann}) generates the same gauge field thermodynamics
as Eq.~(\ref{langevin}), that
it is sufficient to perform a matching calculation of one single  
object to obtain $ \gamma  $. Now it appears that it depends on the choice
of this object whether one obtains contribution from the frequency scale 
$ k ^ 0 \sim C $ or not. 

It is of course possible
that there is a cancellation between various diagrams 
contributing to the $ \vec A \lambda  $ self-energy so that the contribution
discussed above come with a vanishing coefficient, but 
I do not see a reason for
such a cancellation to occur.

\section{Summary and discussion}
\setcounter{equation}0 
\label{sc:summary} 

We have found that the Boltzmann-Langevin equation (\ref{boltzmann}) is not
renormalizable by power counting. There are divergent 1-loop diagrams
for which the action (\ref{action0}) contains no counterterms. Introducing the
necessary counterterms would invalidate the equivalence 
of the  path integral (\ref{path}) and the  
Boltzmann-Langevin equation (\ref{boltzmann}). 
The divergences appear in diagrams with $ A ^ 0 $ and $ \widetilde{W} $
propagators, which for large momenta $ k $ fall off only like $ 1/k $. 
One should emphasize  that these 
divergences are {\em not} related to the ones in classical 
Yang-Mills or in the hard thermal loop effective theory discussed previously
\cite{bms}. Furthermore, they 
do not affect the next-to-leading log order (NLLO) 
calculations.

We have confirmed Arnold and Yaffe's result for the NLLO color conductivity
$ \gamma  $ 
by matching the results for Wilson loops obtained from Eq.~(\ref{boltzmann})  
and Eq.~(\ref{langevin}) using Coulomb gauge. 
We have checked the gauge
fixing independence in  a certain class of flow gauges.  
The dominant contribution is due to the 1-loop $ A ^ 0 $-propagator,
which turned out to be unaffected by loop  momenta 
with $ k ^ 0 \gg g ^ 4 T [\log(1/g)] ^ 3$. 

The gauge field propagator in the theory (\ref{boltzmann}) has a non-trivial
analytic structure for momenta of order $ g T $, which lie outside the
range of validity of (\ref{boltzmann}). The calculation of the
NLLO color conductivity is performed with an ultraviolet
cutoff $ \mu $, $ g ^ 2 T \ll \mu  \ll g T $. The proof that 
Eq.~(\ref{boltzmann}) correctly reproduces the thermodynamics of
$ \vec k \sim g ^ 2 T $ Yang-Mills fields \cite{local} does, however, 
 not assume the presence of such a cutoff which could be a reason for
being concerned.

Finally, we have found that, somewhat surprisingly,  the spatial
gauge field propagator 
for momenta of order $ g ^ 2 T $, unlike the $ A ^ 0 $-propagator, 
 {\em does} receive NLLO  corrections
from loop momenta with $ k ^ 0  
\sim  g ^ 2 T \log(1/g)$.

{\bf Acknowledgments.} 
I would like to thank Poul H.~Damgaard, Edwin Laermann,
Kari Rummukainen and Daniel Zwanziger
 for useful discussions. 
This work was supported by the DFG, grant FOR 339/2-1.


\appendix 
\renewcommand{\theequation}{\thesection.\arabic{equation}}

\section{Properties of $ \ghat $ and related functions}   
\label{ap:ghat}
\setcounter{equation}0

$ \ghat $, defined in Eq.~(\ref{ghat}),
 is symmetric since it is the inverse of a symmetric operator,
\begin{eqnarray}
        \langle \vec v _ 1 | \ghat(k)   |\vec v _ 2 \rangle = 
        \langle \vec v _ 2 | \ghat(k)   |\vec v _ 1 \rangle
        .
\end{eqnarray}
Furthermore, 
\begin{eqnarray}
  \hat{G} (-k) = \hat{G} ^ *(k) 
  .
\end{eqnarray} 
Now multiply
\begin{eqnarray}
        [\chat - i \vhat\cdot k]\ghat(k) = 1
\end{eqnarray}
by $\vbra^0|$ from the left. The collision term drops
out since its $l=0$ eigenvalue vanishes, 
\begin{eqnarray}
        -i \vbra^0| \vhat\cdot k\ghat(k) = \vbra^0|
        .
\end{eqnarray}
Now we use
\begin{eqnarray}
        \vbra^0|\vhat^\mu  &=& \intv{} \vbra \vec v|\vhat^\mu 
        =   \intv{} v^\mu \vbra \vec v|
\nonumber  \\
        &=&    \vbra ^\mu|
\end{eqnarray}
which gives the identity
\begin{eqnarray}
        k ^ i  \vbra^i| \ghat(k) = -i\vbra^0| 
        + k ^ 0 \vbra^0 | \ghat(k) 
        .
\end{eqnarray} 
From this we obtain the following relations
\begin{eqnarray}
        G _ \ell &=& \frac{k ^  0 }{ \vec k ^ 2}        
        \left (k ^ 0  G ^   {00} - i \right )
\\
        k ^ i G ^ {0i} &=& k ^  0  G ^ {00} - i
\end{eqnarray} 
and 
\begin{eqnarray} 
        G ^   {00} &=& \frac{ 1}{k ^ 0} \left ( \frac{ \vec k ^ 2}{k ^ 0}
        G _ \ell + i \right )
\\
        k ^ i G ^ {0i} &=& \frac{ \vec k ^ 2}{k ^ 0} G _ \ell
\end{eqnarray}
When $ k ^ 0 \ll |\vec k| $ we can approximate $ G ^ {0i}(k) $ by
\begin{eqnarray}
        G ^ {0i}(0,\vec k) = -i \frac{ k ^ i}{\vec k ^ 2}
        .
\end{eqnarray} 
The following matrix elements are needed to obtain the results in
Appendix \ref{ap:diagrams} 
\begin{eqnarray}
        \langle ^{ij}|\hat{G} (0, \vec k) |^ l \rangle 
        & = & \frac{ i}{\vec k ^ 2}  
        \left[ c _ 1 G _ {\rm t }(0, \vec k) - \frac{ 1}{d} \right]
        \left[ k ^ i P \tr ^{jl}(\vec k) + k ^ j P \tr ^{il}(\vec k) \right] 
        \label{blP3}
        ,
\\
        k ^ i k ^ j \langle ^{ij}|\hat{G} (0, \vec k) |^ 0 \rangle
        & = & c _ 1 - \frac{ 1}{d} \vec k ^ 2 G ^{00}  (0, \vec k)
        \label{blP4}
        .
\end{eqnarray}

\section{Approximated expressions and power counting for the propagators}
\setcounter{equation}0 
\label{ap:estimates} 

As discussed at the end of Sec.~\ref{sc:nllo}  we will not use the
propagators of Sec.~\ref{sc:pathbl} 
to compute the NLLO color conductivity, but instead 
the approximated expressions 
for momenta small compared to $ m_{\rm D} $, which are much simpler. 
We have to consider two  cases.
The first is  $ k ^ 0, \vec k \sim C $. These are the characteristic momenta
for the $ \widetilde{W} $ fields which perform damped oscillations in such a
 way
that the color current vanishes (see \cite{local}). The second case is
$ k ^ 0 \ll C $ for the characteristic frequencies of the
gauge fields $ k ^ 0 \sim g ^ 4 T $ (modulo logarithms of $ g $). 
There we have to distinguish
$ \vec k \sim  C $ and $ \vec k \ll C $. 

\subsection{$ k ^ 0 \sim \vec k \sim C $ }
When $k _ 0 $ is of order $ C $ we have
$ k _ 0 G \tr \sim 1$. Then  one can neglect $ \vec k^ 2 $ in $ \Delta (k) $,
Eq.~\mref{Delta},
\begin{eqnarray}
        \Delta (k)  = \frac{ 1 }{\mmdebye}
        \left [ \frac{ i }{k _ 0} \frac{ 1}{G \tr } - d \right]
        \times \left( 1 + O(g ^ 2) \right )
\end{eqnarray}
so that 
\begin{eqnarray}
        \,\, \epsfig{file=aaprop.eps,width=1.5cm}\,\, 
  =  \frac{ T}{\mmdebye }P\tr ^{ij} 
        \left[
        \frac{ 1}{k _ 0 ^ 2}
        \frac{ 1}{G \tr (k)} 
        + (k \to -k) 
        \right]
        \times \left( 1 + O(g ^ 2) \right )
        .
\end{eqnarray}
Furthermore, 
\begin{eqnarray}
        |S ^ i \rangle = 
        P\tr^{ij} 
        \frac{ i}{k ^ 0 G\tr}\ghat |^ j \rangle
        \times \left( 1 + O(g ^ 2) \right )
\end{eqnarray} 
and 
\begin{eqnarray}
        \hat{S} \simeq \hat{G} - i \frac{ i }{k ^ 0} \hat{P}_0 - 
        G _\ell ^{-1} \hat{G} |^ i \rangle \hat{k} ^ i \hat{k} ^ j 
        \langle ^ j|
        \hat{G} - G \tr ^{-1} | ^ i \rangle P \tr ^{ij} \langle ^ j
        |\hat{G} 
        \label{blD15}
        .
\end{eqnarray}
To estimate loop diagrams we need order of magnitude
estimates  for the various propagators. From the above expressions one finds
\begin{eqnarray}
        \langle A ^{\mu } \lambda \rangle,   \
        \langle \widetilde{W}  \lambda \rangle
        & \sim & C ^{-1} 
\nonumber 
\\
        \langle A ^\mu  A ^\nu  \rangle, \langle A ^\mu \widetilde{W}  \rangle,
        \langle \widetilde{W}  \widetilde{W}  \rangle
         &\sim& T m^{-2}_{\rm D} C ^{-1}   
        \sim L ^ {-1}   g ^ {-4} T ^{-2}
        \quad (k _ 0, |\vec k| \sim   C)
        \label{estimates1} 
        .
\end{eqnarray} 

\subsection{
$ k ^ 0 \ll C $ }
\label{sc:k0g4t} 
When the frequency $ k ^ 0 $ is small compared to $ C $ the gauge field propagators
are the same as in Ref.~\cite{ay:nllo}. We can approximate
\begin{eqnarray}
        \Delta \simeq \Delta _ 0 
        \equiv \frac{ 1} {\vec k ^ 2 - i \mmdebye k _ 0
                G\tr(0,\vec k)}
        \quad (k _ 0 \sim g ^ 4 T)
        \label{delta0} 
\end{eqnarray}
\begin{eqnarray}
        |S ^ i \rangle = m^2_{\rm D}  P\tr^{ij}
        \Delta _ 0  \hat{G}(0, \vec k) 
        |^ j \rangle
        \quad (k _ 0 \sim g ^ 4 T)
\end{eqnarray}
\begin{eqnarray}
        \langle S ^ 0 (k)| =
        \langle ^ 0| \hat{G} _ 0(\vec k) (1- \hat{P}_0)
\end{eqnarray} 
\begin{eqnarray}
        \,\, \epsfig{file=a0a0prop.eps,width=1.5cm}\,\, 
        = 2 \frac{ T}{\mmdebye }G^{00}(0, \vec k) 
        \times \left( 1 + O(g ^ 2) \right )
        \quad (k _ 0 \sim g ^ 4 T)
        \label{a0a0slow} 
\end{eqnarray}
\begin{eqnarray}        
        \langle \widetilde{W} (k, \vec v) A _ 0 (k ') \rangle
        &=& \nonumber \\ &&  \hspace{-3cm}
        (2 \pi ) ^{d+1} \delta(k + k')
        \frac{T }{m^2_{\rm D} } 
        \langle \vec v | \left ( 1 - \hat{P}_0 - \hat{P}_1 \right )
        \left[\hat{G}(0,\vec k) + \hat{G}(0,-\vec k) \right]|^ 0 \rangle
\end{eqnarray}
\begin{eqnarray}
        \hat{S}(k)  =  \hat{G} _ 0 (\vec k) - \left( \hat{G} _ 0 \hat{P}_0
        + \hat{P}_0 \hat{G} _ 0 \right) + G _  0 ^{00} \hat{P}_0  
        {} + i m^2_{\rm D} k _ 0 \Delta _ 0 \hat{G} _ 0
        |^ i \rangle P \tr ^{ij} \langle ^ j | \hat{G} _ 0
        \label{blD17}
\end{eqnarray}
\subsubsection{$ k _ 0 \sim C ^{3}/m^2_{\rm D} $   and $  \vec k \sim C$}
For these momenta the propagators are of order 
\begin{eqnarray} \nonumber 
        \langle \vec A \lambda \rangle & \sim & m^2_{\rm D} C ^{-3}
\\ \nonumber 
        \langle A ^{0} \lambda \rangle , \langle \widetilde{W} \lambda \rangle
        & \sim & C ^{-1}
\\   \nonumber 
        \langle \vec A \vec A \rangle   &\sim& T m^2_{\rm D} C ^ {-5}
        \sim            L ^ {-5}   g ^ {-8} T ^{-2} 
\\ 
        \langle \vec A \widetilde{W} \rangle &\sim& T C ^{-3}
        \sim    L ^ {-3}   g ^ {-6} T ^{-2} 
\label{estimates2} 
\\ \nonumber 
        \langle A ^{0} A ^{0} \rangle, \langle A ^{0} \widetilde{W}  \rangle,
        \langle \widetilde{W}  \widetilde{W}  \rangle
        &\sim&  T m ^{-2}  _{\rm D} C ^{-1 } 
        \sim L ^ {-1}   g ^ {-4} T ^{-2} 
        .
\end{eqnarray} 

\subsubsection{$  \vec k \sim g ^ 2 T$}
\label{sc:k0g4tl} 
When $ \vec k $ is small compared to $ C $, the propagators for the
gauge fields become the same as in the leading log effective theory
(\ref{langevin}).
We need the expansion of $ \hat{G} (0, \vec k) $ for
$ |\vec k| \ll C $ 
\begin{eqnarray}
        \hat{G} (0, \vec k) & =  & 
         \frac{ d}{c _ 1 \vec k ^ 2} (c _ 1 - i \hat{\vec v} \cdot \vec k)
        \hat{P}_0 (c _ 1 - i \hat{\vec v} \cdot \vec k)
        + (1 - \hat{P}_0) \hat{C} ^{-1}  (1 - \hat{P}_0)
\nonumber \\ &&{} \hspace{-1cm}
        - \frac{ d}{c _ 2 \vec k ^ 2} 
        \left[ \hat{P}_0 (\hat{\vec v} \cdot \vec k)^ 2
        + (\hat{\vec v} \cdot \vec k)^ 2 \hat{P}_0
        \right] 
        + \frac{ d ^ 2}{c _ 2 \vec k ^ 4} 
        \hat{P}_0 (\hat{\vec v} \cdot \vec k)^ 4\hat{P}_0
        + O(\vec k/C ^ 2)
        .
\end{eqnarray} 
In particular, 
\begin{eqnarray}
        \hat{G} (0, \vec k) |^ 0 \rangle = 
        \frac{ d}{\vec k ^ 2} (c _ 1 - i \hat{\vec v} \cdot \vec k)
        |^ 0 \rangle
        + O(C ^{-1} )
        ,
\end{eqnarray}
\begin{eqnarray}
  \langle S ^ 0 (k) | = - i d \frac{ k ^ i}{\vec k ^ 2} \langle ^ i |
  + O(C ^{-1})
  \label{s0th3} 
  ,
\end{eqnarray}  
\begin{eqnarray}
        G ^{00} (0, \vec k) = \frac{ d c _ 1 }{\vec k ^ 2}
        + O(C ^{-1} )
        \label{g00vslow} 
        ,
\end{eqnarray}
and
\begin{eqnarray}
        G \tr(0) = \frac{ 1}{d c_1}
        \label{gt0} 
        .
\end{eqnarray} 
The propagators are of order 
\begin{eqnarray}
        \langle \vec A \lambda \rangle & \sim & m^2_{\rm D} c _ 1 ^{-1}
        (g ^{2}  T ) ^{-2}
\nonumber 
\\
        \langle A ^{0} \lambda \rangle 
        & \sim & (g ^ 2 T )^{-1}
\nonumber 
\\   
        \langle \vec A \vec A \rangle   &\sim &         
         \frac{ T m^2_{\rm D}}{C (g ^ 2 T) ^ 4}
\nonumber 
\\
        \langle A ^ 0 A ^ 0 \rangle &\sim&  L    g ^ {-4} T ^{-2}
                \sim \frac{ T c _ 1}{m^2_{\rm D} (g ^ 2 T) ^ 2}
\nonumber 
\\
        |S ^ i \rangle &\sim&  L ^ {-1}   g ^ {-6} T ^{-2} 
\nonumber 
\\
        \langle A ^ 0  \widetilde{W}  \rangle 
        &\sim&   L ^ 0  g ^ {-4} T ^{-2}
        \sim \frac{ T}{m^2_{\rm D} c _ 1} 
\nonumber 
\\
        \langle \widetilde{W}  \widetilde{W}  \rangle
          &\sim&  \frac{ T}{m^2_{\rm D} c _ 1}  
        \quad \left( k _ 0 \sim \frac{ C (g ^ 2 T)^ 2}{m^2_{\rm D}}
        ,\; \vec k \sim  g ^ 2 T \right) 
      \label{estimates3} 
      .
\end{eqnarray}

\section{Frequency integrals}
\label{ap:frequency}
\setcounter{equation}0
For the frequency integration it is convenient to write $
 \Delta _ 0 $, which is defined in Eq.~(\ref{delta0}), as
\begin{eqnarray}
        \Delta _ 0 ( k) = \frac{ i \Gamma _ {\vec k}}{\vec k ^ 2}
        \frac{ 1}{k ^ 0 + i \Gamma _ {\vec k}}
        ,
\end{eqnarray}
where 
\begin{eqnarray}
        \Gamma _ {\vec k} \equiv \frac{ \vec k ^ 2}{m^2_{\rm D} G \tr 
        (0,\vec k)}
      .
\end{eqnarray}
The frequency integrals for the diagrams in Appendix \ref{ap:diagrams} are
\begin{eqnarray}
        \int _ {k ^ 0} \frac{ 1}{k ^ 0}
        \left[ \Delta _ 0 ( k) - \Delta _ 0 (- k) \right] 
        &=& \frac{ i}{\vec k ^ 2}
        ,
\\
        \int _ {k ^ 0} 
        \Delta _ 0 ( k)  \Delta _ 0 (- k) 
        &=& \frac{ 1}{2 m^2_{\rm D} }
        \frac{ 1}{ \vec k ^ 2 G _{\rm t }(0,\vec k)}
        ,
\\
        \int _ {k ^ 0} \frac{ 1}{k _ 0 ^ 2}
        \left[ \Delta _ 0 ( k) - \Delta _ 0 (- k) \right] ^ 2 
        &=& -m^2_{\rm D} \frac{ G _{\rm t }(0,\vec k)}{\vec k ^ 6}
        .
\end{eqnarray} 
We also need frequency integrals with external momentum $ p ^ \mu =
(0,\vec p) $,
\begin{eqnarray}
        \int _ {k ^ 0} 
        \Delta _ 0 ( k)  \Delta _ 0 (p - k) 
        = \frac{ 1}{m^2_{\rm D} }
        \frac{ 1}{ \vec k ^ 2 G _{\rm t }(0,\vec p - \vec k)
        + (\vec p - \vec k) ^ 2 G _{\rm t }(0,\vec k)}
      ,
\end{eqnarray} 
\begin{eqnarray} 
        \int _ {k ^ 0} 
        \Delta _ 0 ( k) \frac{ 1}{k ^ 0}
        \left[ \Delta _ 0 ( k - p)  - \Delta _ 0 (p - k) \right]  
        && \nonumber \\ && \hspace{-3cm} = 
        \frac{ i G _{\rm t }(0,\vec p - \vec k)}
        {  (\vec p - \vec k) ^ 2
          \left[ \vec k ^ 2 G _{\rm t }(0,\vec p - \vec k)
        + (\vec p - \vec k) ^ 2 G _{\rm t }(0,\vec k) \right] }
    ,
\end{eqnarray} 
\begin{eqnarray} 
        \int _ {k ^ 0} 
        \frac{ 1}{k ^ 0}
        \left[ \Delta _ 0 ( k - p)  - \Delta _ 0 (p - k) \right]  
        = \frac{ i}{(\vec p - \vec k) ^ 2}
        .
\end{eqnarray}

\section{One loop results}   
\label{ap:diagrams}
\setcounter{equation}0
Here we list the Coulomb gauge 
results for the diagrams for the $ \gvec{ \lambda}$$  \gvec{
  \lambda} $ and $ A ^ 0 \gvec{ \lambda }$-selfenergies in the limit of small
external momenta $ p $.

One needs the expansion of $ P   \tr ^{ij} (\vec k - \vec p) $, where 
$ \vec k $
is the loop momentum, for small $ \vec p $,  
\begin{eqnarray}
        P   \tr ^{ij} (\vec k - \vec p) = P   \tr ^{ij} (\vec k)
        + \frac{ p ^ l}{\vec k ^ 2}
        \left[ k ^ i P   \tr ^{jl} (\vec k) + k ^ j P   \tr ^{il} (\vec k) 
        \right] 
        +O(\vec p ^ 2 /\vec k ^ 2)
        .
\end{eqnarray} 

For $ p \to 0 $ the $ \gvec{ \lambda} \gvec{ \lambda} $-selfenergy becomes
 $ \Pi
_{\lambda \lambda } ^ {ij} (p) 
\to \delta ^{ij}\Pi _{\lambda \lambda } (0) $. 
The diagrams contributing to $ - \Pi
_{\lambda \lambda } (0) $ are
\begin{eqnarray} 
  \raisebox{-.45cm}{\epsfig{file=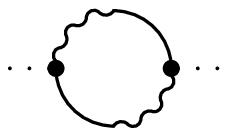,width=2cm}}
        & = & \frac{N g ^ 2 T ^ 2}{m^2_{\rm D} } d(d-1)
        \int _ {\vec k} 
        \frac{ \left(  c _ 1 G \tr(0,\vec k) - \frac{ 1}{d} \right) ^ 2}
        {\vec k ^ 4 G \tr(0,\vec k)}
        \label{d1}
        , 
\\
  \raisebox{-.45cm}{\epsfig{file=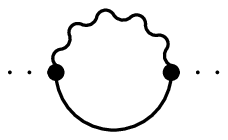,width=2cm}}
        & = & \frac{N g ^ 2 T ^ 2}{m^2_{\rm D} } 
        \int _ {\vec k} 
        \Bigg\{ - \frac{2 d }{\vec k ^ 2}P\tr ^{lm} 
          \langle ^{kl} | \hat{G} (0,\vec k)|^{km} \rangle   
\nonumber \\ & & {}\hspace{2cm}        
        - d(d-1) 
        \frac{ \left(  c _ 1 G \tr(0,\vec k) - \frac{ 1}{d} \right) ^ 2}
        {\vec k ^ 4 G \tr(0,\vec k)}
        \Bigg\}
        \label{d2}
        , 
\\
  \raisebox{-.45cm}{\epsfig{file=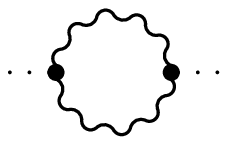,width=2cm}}
        & = & -2 c _ 1 ^ 2 d(d-1) \frac{N g ^ 2 T ^ 2}{m^2_{\rm D} }
        \int _ {\vec k} 
        \frac{ G \tr(0,\vec k)}{\vec k ^ 4}
        \label{d3}
        , 
\\
  \raisebox{-.45cm}{\epsfig{file=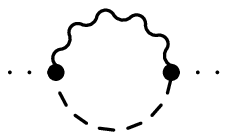,width=2cm}}
        & = & - \frac{2( d-1)}{d} \frac{N g ^ 2 T ^ 2}{m^2_{\rm D} } 
        \int _ {\vec k} 
        \frac{ G ^{00}(0,\vec k)}{\vec k ^ 2} 
        \label{d4} 
        ,
\\
  \raisebox{-.45cm}{\epsfig{file=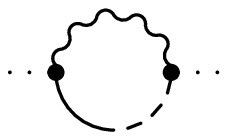,width=2cm}}
        & = & 2 \frac{N g ^ 2 T ^ 2}{m^2_{\rm D} } 
        \int _ {\vec k} \frac{ 1}{\vec k ^ 4}
        \left( c _ 1 -\frac{ 1}{d}\vec k ^ 2 G ^{00} (0,\vec k) \right) 
        \label{pg23.2}
        ,
\\
  \raisebox{-.45cm}{\epsfig{file=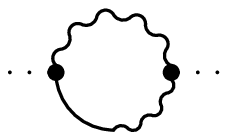,width=2cm}}
        & = & 0
        \label{pg22.1}
        .
\end{eqnarray}
Note that the  diagram (\ref{pg23.2}) enters $ \Pi  _{\lambda  \lambda  } $
with an additional
factor 2 because the reflected diagram gives the same
contribution.  

Arnold and Yaffe \cite{ay:nllo} write their results 
in terms of  a different matrix element than the one Eq.~(\ref{d2}).
Ours is related to theirs by
\begin{eqnarray} 
  P\tr ^{lm} 
  \langle ^{kl} | \hat{G} (0,\vec k)|^{km} \rangle 
  =  P\tr ^{lm}       \langle ^0 | \hat{v} ^ k \hat{v} ^ l 
  \hat{G} (0,\vec k) \hat{v} ^ k \hat{v} ^ m |^0 \rangle
  + \frac{ 2 c _ 1}{d \vec k ^ 2} - \frac{ d + 1}{d ^ 2} G ^{00} (0, \vec k)
  \label{blN29} 
  .
\end{eqnarray}
With this relation the sum of Eqs.~(\ref{d1})--(\ref{pg23.2}) 
gives
\begin{eqnarray}
  \Pi
  _{\lambda \lambda } (0) =  
  2d \frac{N g ^ 2 T ^ 2}{m^2_{\rm D} } 
        \int _ {\vec k} \left\{ \frac{ 1}{\vec k ^ 2}
          P\tr ^{lm}   (\vec k)    \langle ^0 | \hat{v} ^ k \hat{v} ^ l 
  \hat{G} (0,\vec k) \hat{v} ^ k \hat{v} ^ m |^0 \rangle
   +(d-1) c _ 1 ^ 2 \frac{ G\tr (0, \vec k)}
  {\vec k ^ 4} \right\} 
  \label{blN28.1}
  .
\end{eqnarray}

The $ \gvec \lambda  A ^ 0 $-selfenergy $ \Pi  _{\lambda  A ^ 0}^ i (p)$ 
vanishes for $ p = 0 $. Expanding around  $ p =0 $ 
the lowest order contribution
in $ \vec p $ is given by the diagrams
\begin{eqnarray} 
  \raisebox{-.45cm}{\epsfig{file=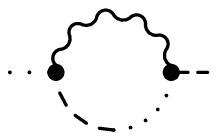,width=2cm}}
        & = & N g ^ 2 T \frac{ d-1}{d} p ^ i 
        \int _ {\vec k} \frac{ 1}{\vec k ^ 4}
\end{eqnarray} 
\begin{eqnarray} 
  \raisebox{-.5cm}{\epsfig{file=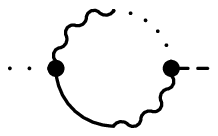,width=2cm}}
       \quad + \quad
  \raisebox{-.5cm}{\epsfig{file=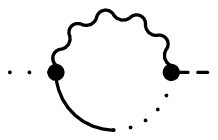,width=2cm}}
\nonumber &=& \\ 
       &&\hspace{-2cm}    - N g ^ 2 T (d-1) p ^ i 
        \int _ {\vec k} \frac{ 1}{\vec k ^ 4}
        \left( c _ 1 G \tr (0,\vec k) - \frac{ 1}{d} \right) 
        \label{pg25.3}
\end{eqnarray} 
\begin{eqnarray} 
  \raisebox{-.5cm}{\epsfig{file=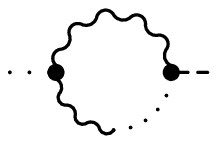,width=2cm}}
        & = & - N g ^ 2 T c _ 1 (d -1) 
        p ^ i 
        \int _ {\vec k} \frac{ 1}{\vec k ^ 4}
         G \tr (0,\vec k) 
        \label{pg26.3}
\end{eqnarray}
the sum of which gives  
\begin{eqnarray} 
  - \Pi _{\lambda A ^ 0} ^ i = 
  \raisebox{-.45cm}{\epsfig{file=lama0blob.eps,width=2cm}}
        = 2 N g ^ 2 T (d - 1) p ^ i \int _{\vec k} \frac{ 1}{\vec k ^ 4}
        \left( \frac{ 1}{d} 
          - c _ 1 G\tr (0, \vec k) \right) 
        \label{bl31}
        .
\end{eqnarray}


\end{document}